\documentclass[a4paper,11pt,usenames,dvipsnames]{article}
\usepackage{jheppub}

\usepackage{epsfig,multicol}
\usepackage{amsmath}
\usepackage{graphics}
\usepackage{graphicx}
\usepackage{dcolumn}
\usepackage{amssymb}
\usepackage{amsthm}
\usepackage{amsfonts}
\usepackage{subfigure}
\usepackage{setspace}

\title{Dissipative dynamics of an open quantum battery in the BTZ spacetime}

\author[a]{Zehua Tian, \footnote{Corresponding author, Email:tzh@hznu.edu.cn}}
\author[b]{Xiaobao Liu,}
\author[c] {Jieci Wang,}
\author[c] {Jiliang Jing}
\affiliation[a]{School of Physics, Hangzhou Normal University, Hangzhou, Zhejiang 311121, China}
\affiliation[b]{Department of physics and electrical engineering, Liupanshui Normal University, Liupanshui 553004, Guizhou, China}
\affiliation[c]{ Department of
Physics, Key Laboratory of Low Dimensional Quantum Structures and
Quantum Control of Ministry of Education, and Synergetic Innovation
Center for Quantum Effects and Applications, Hunan Normal
University, Changsha, Hunan 410081, China}
\emailAdd{tzh@hznu.edu.cn}
\emailAdd{xiaobaoliu@hotmail.com}
\emailAdd{ jcwang@hunnu.edu.cn}
\emailAdd{jljing@hunnu.edu.cn}

\abstract{ We consider how charging performances of a quantum battery, modeled as a two-level system, are influenced by the presence of vacuum fluctuations of a quantum field satisfying the Dirichlet, transparent, and Neumann boundary conditions in the BTZ spacetime. The quantum battery is subjected to an external static driving which works as a charger. Meanwhile, the quantum field is assumed to be coupled to both longitudinal and transverse spin components of the quantum battery including decoherence and pure dephasing mechanisms. Charging and discharging dynamics of the quantum battery are derived by extending the previous open quantum system approach in the relativistic framework to this more general scenario including both the driving and multiple coupling. Analytic expressions for the time evolution of the energy stored are presented. We find that when the driving amplitude is stronger/weaker than the energy-level spacing of the quantum battery the pure dephasing dissipative coupling results in better/worse charging performances than the decoherence dissipative coupling case. We also find that higher local Hawking temperature helps to improve the charging performance under certain conditions compared with the closed quantum battery case, implying the feasibility of energy extraction from vacuum fluctuations in curved spacetime via dissipation in charging protocol. Different boundary conditions for quantum field may lead to different charging performance. Furthermore, we also address the charging stability by monitoring the energy behaviour after the charging protocol has been switched off. Our study presents a general framework to investigate relaxation effects in curved spacetime, and reveals how spacetime properties and field boundary condition affect the charging process, which in turn may shed light on the exploration of the spacetime properties and thermodynamics via the charging protocol.
}

\begin{document}
\maketitle

\section{Introduction}
In the era of quantum supremacy for quantum computing \cite{nielsen2010quantum, QS}, the  so called ``quantum thermodynamics" \cite{ RevModPhys.81.1665, RevModPhys.83.771, RevModPhys.93.035008, RevModPhys.93.041001} has emerged  
to explore the thermodynamics concepts, such as work and heat, at the quantum scale, as well as to develop the new thermodynamic protocols
exploiting quantum mechanical resources in the thermodynamic protocol, such as entanglement and coherence, to outperform their classical counterparts. 
A simple paradigm to explore this issue is a quantum battery (QB): a designed device operating in the quantum regime capable of temporarily storing energy for later use \cite{PhysRevE.87.042123, Campaioli2018, Campaioli:2023ndh}.

The notion of QB was first introduced in the seminal work \cite{PhysRevE.87.042123} by Alicki and Fannes, showing entangling unitary controls extract in general more work than independent ones. Afterward, the development of different figures of merit, such as the charging time, the associated power, and the 
amount of extractable work, has been a major focus on the QB physics \cite{Campaioli:2023ndh, Binder_2015, PhysRevLett.122.047702, PhysRevLett.111.240401, PhysRevLett.111.050601, PhysRevLett.131.030402, PhysRevLett.129.130602, Fischer}. To improve the charging performance, a lot of methods have been developed to assist the charging protocol, including collective quantum resources \cite{PhysRevLett.120.117702, PhysRevLett.118.150601, PhysRevLett.128.140501}, 
ordered and disordered interactions \cite{PhysRevA.101.032115}, external environment engineering \cite{PhysRevA.107.032203, PhysRevE.104.064143, PhysRevA.109.022226, PhysRevA.102.052223}, collisional model \cite{PhysRevLett.127.100601, PhysRevResearch.5.013155}, Sachdev-Ye-Kitaev model \cite{PhysRevLett.125.236402, SYK} and so on; see Ref. \cite{Campaioli:2023ndh} for an extensive review and a comprehensive list of references. Beyond the theoretical analysis, the experimental implementation of various kinds of QBs and different charging protocols has also been reported recently \cite{Campaioli:2023ndh, PhysRevB.108.L180301, PhysRevLett.131.240401, Hu_2022, PhysRevA.106.042601, batteries8050043}.
Furthermore, since any systems are unavoidably coupled to the environment, at least subjected to vacuum fluctuations at the quantum scale, relaxation and dephasing would play a key role in the system dynamics. In this sense, the QB's figures of merit, including 
the maximum average energy stored during the charging protocol, the charging power,
the charging stability and so on, inevitably undergoes the relaxation and dephasing, which may even dominate the QB charging performance. It is thus of great importance to learn the charging performance in the presence of inevitable external environmental action \cite{Campaioli:2023ndh, PhysRevB.99.035421, PhysRevA.102.060201, PhysRevA.103.033715, Carrega_2020, PhysRevLett.122.210601, PhysRevE.104.044116, PhysRevA.104.032207}.

On the other hand, open quantum system approach \cite{breuer2002theory, 10.1063/1.5115323, PhysRevLett.91.070402} has been extended to the relativistic framework by considering a two-level quantum system (or detector) coupled to vacuum fluctuations of quantum fields in different spacetimes and undergoing various relativistic motions \cite{PhysRevA.70.012112, PhysRevA.106.062440, PhysRevD.77.024031, PhysRevLett.106.061101, Dsopen, PhysRevA.107.012208}. 
Because of the quantum field scattering by spacetime, through the interaction between the detector and fields the information of spacetime properties and relativistic motions may be encoded into the dynamical evolution of the detector. Therefore, the open quantum system approach has been fruitfully applied in the relativistic framework to exploring various physics, such as understanding and detecting relativistic effects and spacetime properties \cite{ PhysRevD.77.024031, PhysRevLett.106.061101, PhysRevA.85.032105, Tian1, Liu, FIDs, tian2, FENG2022136992, FIBH, PhysRevLett.129.160401, tian3}, measuring quantum state and work distributions of quantum fields \cite{tian4, PhysRevA.102.052219, PhysRevLett.122.240604}, and so on. Furthermore, this simple but effective detector-field model has been extensively utilized in the exploration of interplay between relativity and quantum mechanics \cite{PhysRevResearch.3.043056, PhysRevLett.129.181301, PhysRevD.102.045002, PhysRevD.108.L101702, PhysRevLett.125.131602}.

In this paper, we extend the charging protocol of QB to the relativistic framework, trying to explore thermodynamics, relativity and the interplay with the laws of quantum mechanics jointly through dissipation dynamics. We therefore inspect the dynamics of a QB, modeled as a two-level atom. It is coupled to the vacuum fluctuations of quantum field in the BTZ spacetime responsible for dissipation, and meanwhile is operated by a static external classical driving acting as a charger.
Note that the external driving on quantum system is essential to quantum state preparation, system manipulation and final state readout in the quantum information process \cite{Statep1, doi:10.1126/science.1149858, RevModPhys.93.025001, RevModPhys.91.045001}. To obtain the QB's dynamics, we extend the previous open quantum system approach \footnote{Usually, the system is assumed to be coupled to quantum field only with transverse component and without external driving.} in the relativistic framework to a more general and practical scenario where not only are both the QB's transverse and longitudinal spin components coupled to the quantum field, but also is the external driving considered to act on the system.
We derive the analytical expression of the average energy stored in the QB and analyze how its dynamics is affected by the driving amplitude, BTZ spacetime properties, the boundary conditions on quantum field, and different dissipative couplings. We also investigate the corresponding charging stability in these scenarios after switching off the charging. Interestingly, it is found that which kind of dissipative coupling, decoherence or pure dephasing, leads to a better charging performance depends on the weight between the driving amplitude and  the energy-level spacing of the QB. Besides, Hawking radiation may assist on the charging protocol under certain conditions, manifesting the improvement of the maximum average energy stored in the QB compared with the closed one. It means that one can extract energy from vacuum fluctuations in curved spacetime via dissipation in charging protocol.

Note that the BTZ spacetime is an exact solution of the Einstein equation in $(2+1)$-dimensional gravity that is found by M\'aximo Ba\~nados, Claudio Teitelboim, and Jorge Zanelli (BTZ) \cite{PhysRevLett.69.1849}. As a result of that, the BTZ solution might provide a concrete paradigm for general relativity in $(2+1)$ dimensions to explore the foundations of classical and quantum gravity \cite{DESER1984220, CQG}, and thus has attracted a lot attention. It has been found that 
the BTZ black hole has an event horizon and (in the rotating case) an inner horizon, appears as the final state of collapsing matter, and shares the thermodynamic properties much like those of $(3+1)$-dimensional black hole. However, it is asymptotically anti-de Sitter rather than asymptotically flat, and has no curvature singularity at the origin, which are quite different from the Schwarzschild and Kerr solutions \cite{PhysRevD.49.1929, carlip19952}. Furthermore, in the BTZ spacetime,
the Wightman functions for a conformally coupled scalar field (in the Hartle-Hawking vacuum) are known analytically \cite{PhysRevD.49.1929, PhysRevD.59.104017}.
Until now, many interesting quantum phenomena associated with the BTZ spacetime has been explored, e.g., the response of the Unruh-DeWitt particle detectors \cite{PhysRevD.86.064031}, quantum fluctuations \cite{POURHASSAN2017325}, correlation harvesting \cite{Henderson_2018, PhysRevD.106.025010}, anti-Hawking effect \cite{CAMPOS2021136198, HENDERSON2020135732, PhysRevD.106.045018}, quantum signatures of black hole mass superpositions \cite{PhysRevLett.129.181301}, Fisher information \cite{FIBH}, Lamb shift \cite{Lamb},  holographic complexity \cite{HC} and so on.  As a further step, 
we are going to investigate, in the present paper, the dissipative dynamics of an open QB coupled with the fluctuating conformal massless scalar field in vacuum in the BTZ spacetime, hoping to further understand the properties of the BTZ spacetime in terms of charging protocol, as well as the dissipative dynamics of open
QB in the relativistic framework. 

The outline of this paper is as follows. In section \ref{section2} we present the model of a single cell QB coupled to a vacuum fluctuations of quantum field. The charging performance indicators, such as the average energy stored in the QB, are calculated here.  In section \ref{section3} we consider the dynamics of the average energy stored when the open QB is located in the BTZ spacetime, and explore how the driving amplitude, BTZ spacetime properties, the boundary conditions on quantum field, and different dissipative couplings affect the charging performance, including both the dynamics of the average energy stored and the charging stability. In section \ref{section4}, we study the quantum battery physics in AdS spacetime and compare the results with that of the BTZ spacetime case.
Finally, discussions and summary of the main results are present in section \ref{section5}.

\section{The model} \label{section2}
In this section we will introduce our open quantum battery model and the relevant physical quantities to characterize the charging and discharging performances of a quantum battery. In our paper, $\hbar=1$ is set, $\sigma_k~(k=x, y, z)$ indicates the usual $k$-th Pauli matrix.

\subsection{Open quantum battery model}
A single cell QB is modeled as a two-level system with the following Hamiltonian,
\begin{eqnarray}
H_\text{QB}=\frac{\Delta}{2}\sigma_z,
\end{eqnarray}
where $\Delta$ denotes the energy-level spacing between the system's ground and excited state, i.e., $|g\rangle$ and $|e\rangle$, which can be seen as the empty and the full cell configuration, respectively. Note that we will consider all the physics in the framework of the QB which has the proper time $\tau$. In order to explore the charging protocol, at time $\tau=0^{+}$ we switch on the charger which is described by an external classical field $A$ coupled with the $\sigma_x$ component of the QB \cite{Binder_2015, PhysRevLett.118.150601, PhysRevLett.120.117702, Campaioli:2023ndh},
\begin{eqnarray}
H_\text{C}=-\frac{A}{2}\sigma_x.
\end{eqnarray}
In addition to that, the QB is assumed to be locally coupled to a fluctuating massless scalar field $\phi(x(\tau))$ in vacuum, which may result in dissipation of the QB. The field's Hamiltonian reads
\begin{eqnarray}
H_\text{F}(\tau)=\int d^3k\omega_\mathbf{k}a^\dagger_\mathbf{k}a_\mathbf{k}\frac{dt}{d\tau},
\end{eqnarray}
where $a_\mathbf{k}^\dagger$, $a_\mathbf{k}$ are the creation and annihilation operators of the scalar field. The specific QB-field coupling is assumed to be linear along both the longitudinal $(z)$ and transverse $(x)$ directions, 
\begin{eqnarray}\label{Interaction-H}
H_\text{I}=\mu(\sin\theta\sigma_z+\cos\theta\sigma_x)\phi(x(\tau)),
\end{eqnarray}
where $\mu$ is a small coupling constant. 
$\theta$ here is to control the weight of the QB-field coupling that along the longitudinal (z) and transverse (x) directions. This interaction \eqref{Interaction-H} captures both decoherence $(\theta=0)$ and pure dephasing $(\theta=\pi/2)$ processes. 
Note that in additional to the linear coupling between matter and field \eqref{Interaction-H} which is commonly used in study of Unruh-DeWitt detector when learning the quantum field theory in curved spacetime, actually other various coupling forms, e.g.,  the derivative coupling \cite{K-J-Hinton_1983, P-G-Grove_1986, Juarez-Aubry_2014, Louko, Jorma, PhysRevD.98.065006, Erickson, PhysRevD.104.025001, Mann}, and quadratic coupling \cite{K-J-Hinton_1984, Norikazu-Suzuki_1997, doi:10.1142/S0217732302007545, PhysRevD.93.024019, PhysRevD.96.085012, PhysRevD.103.125021}, have also absorbed much attention. Besides, weak coupling is common for matter-field interaction in nature, while its strong and even ultrastrong coupling regime have been demonstrated in a variety of settings recently \cite{RevModPhys.90.021002, RevModPhys.91.025005, Kockum, RevModPhys.92.025003}.

To study the dynamical evolution of the QB, we start by considering an unitary rotation in the spin space $R(\Theta)=e^{-i\frac{\Theta}{2}\sigma_y}$, with the angle $\Theta$ chosen in such a way to project the QB-Charger part $H_\text{QB}+H_\text{C}$ only along the $z$ axis. After this rotation, we can find 
\begin{eqnarray}\label{HS}
\nonumber
\tilde{H}_\text{BQ}&=&RH_\text{QB}R^{\dagger}=\frac{\Delta}{2}(\cos\Theta\sigma_z+\sin\Theta\sigma_x),
\\       \nonumber
\tilde{H}_\text{C}&=&RH_\text{C}R^{\dagger}=\frac{A}{2}(\sin\Theta\sigma_z-\cos\Theta\sigma_x),
\\
\tilde{H}_\text{I}&=&RH_\text{I}R^{\dagger}=\mu(\sin(\theta-\Theta)\sigma_z+\cos(\theta-\Theta)\sigma_x)\phi(x(\tau)).
\end{eqnarray}  
Choosing $\cos\Theta=\Delta/\sqrt{\Delta^2+A^2}$ and $\sin\Theta=A/\sqrt{\Delta^2+A^2}$, then $\tilde{H}_\text{BQ}+\tilde{H}_\text{C}=\frac{1}{2}\sqrt{\Delta^2+A^2}\sigma_z$. In the beginning, we assume the total state of the QB and the quantum field
is prepared at  $\rho_\text{QB}(0)\otimes|0\rangle\langle0|$, where $\rho_\text{QB}(0)=\frac{1}{2}(\mathrm{I}+\vec{\mathbf{w}}(0)\cdot\vec{\sigma})$ is initial reduced density matrix of the QB, and $|0\rangle$ is the vacuum of the scalar field, defined by $a_\mathbf{k}|0\rangle=0$ for all $\mathbf{k}$. Therefore, the rotated reduced initial matrix of the QB is $\tilde{\rho}_\text{QB}(0)=R(\Theta)\rho_\text{QB}(0)R^\dagger(\Theta)=\frac{1}{2}(\mathrm{I}+\vec{\mathbf{r}}(0)\cdot\vec{\sigma})$, satisfying
\begin{eqnarray}
\nonumber
r_1(0)&=&w_1(0)\cos\Theta+w_3(0)\sin\Theta,
\\     \nonumber
r_2(0)&=&w_2(0),
\\
r_3(0)&=&w_3(0)\cos\Theta-w_1(0)\sin\Theta.
\end{eqnarray}

For the whole system, its equation of motion in the interaction picture yields
\begin{eqnarray}\label{EM}
\frac{\partial\tilde{\rho}_\text{tot}(\tau)}{\partial\tau}=-i[\tilde{H}_\text{I}(\tau), \tilde{\rho}_\text{tot}(\tau)].
\end{eqnarray}
Since we are interested in the dynamics of the QB, we trace over the degree of freedom of the field. Besides, in the limit of weak coupling between the QB and the field we perform the Born approximation and the Markov approximation  \cite{breuer2002theory}. From Eq. \eqref{EM} we can finally find the evolution of the reduced density matrix 
$\tilde{\rho}_\text{QB}(\tau)$ can be written in the Lindblad master equation form \cite{breuer2002theory, 10.1063/1.5115323},
\begin{eqnarray}\label{LME}
\frac{\partial\tilde{\rho}_\text{QB}(\tau)}{\partial\tau}=-i[\tilde{H}_\text{eff}, \tilde{\rho}_\text{QB}(\tau)]+\mathcal{L}[\tilde{\rho}_\text{QB}(\tau)],
\end{eqnarray}
where 
\begin{eqnarray}
\mathcal{L}[\tilde{\rho}]=\sum_{i=\pm, z}[L_i\tilde{\rho}\,L^\dagger_i-\frac{1}{2}(L^\dagger_iL_i\tilde{\rho}+\tilde{\rho}\,L^\dagger_iL_i)],
\\
\tilde{H}_\text{eff}=\frac{1}{2}\Omega\sigma_z=\frac{1}{2}\{\sqrt{\Delta^2+A^2}+\mathrm{Im}(\Gamma_++\Gamma_-)\}\sigma_z.
\end{eqnarray}
Note $\tilde{H}_\text{eff}$ is the effective Hamiltonian in which $\Omega$ denotes the effective energy level-spacing of the QB with correction term $\mathrm{Im}(\Gamma_++\Gamma_-)$ being the Lamb shift. Usually the Lamb shift can be neglected because it is far less than $\sqrt{\Delta^2+A^2}$, i.e., $\Omega\approx\sqrt{\Delta^2+A^2}$. Besides, $L_+=\sqrt{\gamma_+}\sigma_+$, $L_-=\sqrt{\gamma_-}\sigma_-$, $L_z=\sqrt{\gamma_z}\sigma_z$
with $\sigma_\pm=\frac{1}{2}(\sigma_x\pm\,i\sigma_y)$ and 
\begin{eqnarray}\label{SP}
\nonumber
\gamma_\pm&=&2\mathrm{Re}\Gamma_\pm=\mu^2\cos^2(\theta-\Theta)\int^{+\infty}_{-\infty}d\Delta\tau\,e^{\mp\,i\Omega\Delta\tau}G(x(\tau), x(\tau^\prime)),
\\ 
\gamma_z&=&\mu^2\sin^2(\theta-\Theta)\int^{+\infty}_{-\infty}d\Delta\tau\,G(x(\tau), x(\tau^\prime)),
\end{eqnarray}
where $G(x(\tau), x(\tau^\prime))$ is the field Wightman function and $\Delta\tau=\tau-\tau^\prime$.

We assume the time evolution density matrix of the QB can be written as  
\begin{eqnarray}\label{TES}
\tilde{\rho}_\text{QB}(\tau)=\frac{1}{2}(\mathrm{I}+\vec{\mathbf{r}}(\tau)\cdot\vec{\sigma}),
\end{eqnarray}
so the corresponding initial state is $\tilde{\rho}_\text{QB}(0)=\frac{1}{2}(\mathrm{I}+\vec{\mathbf{r}}(0)\cdot\vec{\sigma})$. Substituting the density matrix $\tilde{\rho}_\text{QB}(\tau)$ 
 \eqref{TES} into the master equation \eqref{LME}, the time dependent state parameters $\vec{\mathbf{r}}(\tau)$ after a series of calculations are found to be 
\begin{eqnarray}\label{Rotation-S}
\nonumber
r_1(\tau)&=&r_1(0)\cos(\Omega\tau)e^{-\frac{1}{2}(\gamma_++\gamma_-+4\gamma_z)\tau}-r_2(0)\sin(\Omega\tau)e^{-\frac{1}{2}(\gamma_++\gamma_-+4\gamma_z)\tau},
\\       \nonumber
r_2(\tau)&=&r_1(0)\sin(\Omega\tau)e^{-\frac{1}{2}(\gamma_++\gamma_-+4\gamma_z)\tau}+r_2(0)\cos(\Omega\tau)e^{-\frac{1}{2}(\gamma_++\gamma_-+4\gamma_z)\tau},
\\
r_3(\tau)&=&r_3(0)e^{-(\gamma_++\gamma_-)\tau}+\frac{\gamma_+-\gamma_-}{\gamma_++\gamma_-}(1-e^{-(\gamma_++\gamma_-)\tau}).
\end{eqnarray}
Rotating $\tilde{\rho}_\text{BQ}(\tau)$ in Eq. \eqref{TES} with state parameters \eqref{Rotation-S}, we can obtain the evolving state of the QB, $\rho_\text{QB}(\tau)=R^\dagger(\Theta)\tilde{\rho}_\text{QB}(\tau)R(\Theta)$. Note that the unavoidable coupling with an environment is responsible for dissipation, leading to relaxation and dephasing of the QB during the charging protocol, which are respectively characterized by the incoherent relaxation rate $\Gamma^\text{r}_\theta=\gamma_++\gamma_-$ and the dephasing rate $\Gamma_\theta=\frac{1}{2}(\gamma_++\gamma_-+4\gamma_z)$ shown in Eq. \eqref{Rotation-S}.  During the relaxation process, the population of the quantum state of the QB (the diagonal terms in the density matrix of quantum state) might vary with respect to the evolution time. However, the dephasing process would not change the population of the quantum state, but lead to the degradation of coherence (reducing the off-diagonal terms in the density matrix of quantum state). As a result of that, the environment-induced dissipation can inevitably affect the dynamics of the QB, thus the charging performance in terms of the figures of merit of the QB.

\subsection{Charging performance indicators}
To characterize the charging performance, we first need to clarify the time evolution of the spin components $\langle\sigma_k(\tau)\rangle~(k=x, y, z)$, which 
can be written as averages over the time dependent total density matrix, $\rho_\text{tot}(\tau)$ driven by the total Hamiltonian $H=H_\text{QB}+H_\text{C}+H_\text{F}+H_\text{I}$
\begin{eqnarray}
\langle\sigma_k(\tau)\rangle=\mathrm{Tr}[\rho_\text{tot}(\tau)\sigma_k].
\end{eqnarray}
These averages can be represented in terms of the time dependent reduced density matrix $\rho(\tau)$ as
\begin{eqnarray}
\langle\sigma_k(\tau)\rangle=\mathrm{Tr}[\rho(\tau)\sigma_k],
\end{eqnarray}
where $\rho(\tau)=\mathrm{Tr}_\text{F}[\rho_\text{tot}(\tau)]$, meaning that the degree of freedom of the field has been traced over.

Energy exchanges between the different subparts, and thus charging dynamics, at time $\tau$ can be then determined by studying the energy variations
\begin{eqnarray}\label{A-E}
\langle\,E_\text{s}(\tau)\rangle=\langle\,H_\text{s}(\tau)\rangle-\langle\,H_\text{s}(0)\rangle,
\end{eqnarray}
where $\text{s}=\text{QB, C, FI}$. Note that $\text{FI}$ denotes the field plus the interaction between the field and the QB. These energy variations denote the energy change of $\text{s}$-system during the time interval $\tau$. When $\langle E_\text{s}(\tau)\rangle>0$, it means the energy is injected into the s-system. While for $\langle E_\text{s}(\tau)\rangle<0$, it means the energy flows out of the s-system. For example, for the QB system,
$\langle E_\text{QB}(\tau)\rangle>0$ means the QB is charged during charging process and the energy is deposited in which for later use. While $\langle E_\text{QB}(\tau)\rangle<0$ means the QB is discharged and the energy flows out of which for external utilization.

Recall that for $\tau>0$, 
since we have assumed that the driving is static, we have $\dot{H}=0$, with an energy balance of the form 
\begin{eqnarray}
\langle\,E_\text{QB}(\tau)\rangle+\langle\,E_\text{C}(\tau)\rangle+\langle\,E_\text{FI}(\tau)\rangle=0.
\end{eqnarray}
Note that this condition must be hold for any driving amplitude and dissipation.

From now on, we calculate the average energy of the QB, $\langle\,H_\text{QB}(\tau)\rangle$, and that of the changer, $\langle\,H_\text{C}(\tau)\rangle$.
Using the rotation $R(\Theta)$ and the expressions \eqref{HS}, we can directly obtain   
\begin{eqnarray}\label{H-E}
\nonumber
\langle\,H_\text{QB}(\tau)\rangle&=&\mathrm{Tr}[H_\text{QB}\rho_\text{QB}(\tau)]=\mathrm{Tr}[R(\Theta)H_\text{QB}R^\dagger(\Theta)R(\Theta)\rho_\text{QB}
(\tau)R^\dagger(\Theta)]
\\
&=&\mathrm{Tr}[\tilde{H}_\text{QB}\tilde{\rho}_\text{QB}(\tau)]=\frac{\Delta}{2}\bigg[\langle\sigma_z(\tau)\rangle\cos\Theta+\langle\sigma_x(\tau)\rangle\sin\Theta\bigg],
\\  \nonumber
\langle\,H_\text{C}(\tau)\rangle&=&\mathrm{Tr}[H_\text{C}\rho_\text{C}(\tau)]=\mathrm{Tr}[R(\Theta)H_\text{C}R^\dagger(\Theta)R(\Theta)\rho_\text{C}
(\tau)R^\dagger(\Theta)]
\\
&=&\mathrm{Tr}[\tilde{H}_\text{C}\tilde{\rho}_\text{C}(\tau)]=\frac{A}{2}\bigg[\langle\sigma_z(\tau)\rangle\sin\Theta-\langle\sigma_x(\tau)\rangle\cos\Theta\bigg].
\end{eqnarray}
We can further calculate the above average energies using the solutions shown in \eqref{Rotation-S}. It is given by
\begin{eqnarray}\label{H-E2}
\nonumber
\langle\,H_\text{QB}(\tau)\rangle&=&\frac{\Delta}{2}\bigg[r_3(\tau)\cos\Theta+r_1(\tau)\sin\Theta\bigg],
\\  
\langle\,H_\text{C}(\tau)\rangle&=&\frac{A}{2}\bigg[r_3(\tau)\sin\Theta-r_1(\tau)\cos\Theta\bigg].
\end{eqnarray}
Note that $r_3(\tau)$ and $r_1(\tau)$ are respectively related to the diagonal and off-diagonal terms of the density matrix of the QB's state, whose dynamics 
are determined by the incoherent relaxation rate and the dephasing rate respectively. Besides, both the incoherent relaxation rate $\Gamma^\text{r}_\theta$ and the dephasing rate $\Gamma_\theta$ depend on the quantum field correlation function shown in \eqref{SP}, thus $r_3(\tau)$ and $r_1(\tau)$ depend on the spacetime properties due to the scattering off quantum field. Therefore, one can expect that the charging dynamics depends on the spacetime properties, and in turn one can also learn the spacetime properties from the charging physics.

\section{Quantum battery in BTZ spacetime} \label{section3}

In this section, we consider the QB which is coupled to conformal massless scalar fields in vacuum in the BTZ spacetime. The corresponding line element of the BTZ spacetime can be written in Schwarzschild-like coordinates \cite{PhysRevLett.69.1849, PhysRevD.49.1929, S-Carlip-1995} as
\begin{eqnarray}
ds^2=-\frac{r^2-M\ell^2}{\ell^2}dt^2+\frac{\ell^2}{r^2-M\ell^2}dr^2+r^2d\phi^2.
\end{eqnarray}
We can see from the above metric that, unlike the Schwarzschild black hole case, the BTZ black hole  is asymptotically anti-de Sitter
with a negative cosmological constant, $\Lambda=-1/\ell^2$. Besides, it has a horizon at $r_\text{h}=\sqrt{M}\ell$
with $M$ denoting the mass of the BTZ solution.

To obtain the charging dynamics of the QB in the BTZ spacetime, we first need to know the Wightman function of the conformal scalar field which may capture the character of the quantum field vacuum fluctuations and plays an important role in the dynamics of the QB through the QB-field interaction. In this paper,
we assume that the scalar field is in the Hartle-Hawking vacuum in the BTZ spacetime. It is interesting to note that the BTZ spacetime can be obtained 
by a topological identification of $\mathrm{AdS}_3$ spacetime. Therefore, using the method of images,
the corresponding Wightman function in the BTZ spacetime
for the conformal massless scalar field in the Hartle-Hawking vacuum can be obtained in terms of the corresponding Wightman function in $\mathrm{AdS}_3$ spacetime \footnote{Go beyond the standard quantization of quantum field, an alternative quantization in AdS/CFT  has been introduced in Ref. \cite{AdsCFT}.} \cite{PhysRevD.49.1929},
\begin{eqnarray}\label{Wightman-F}
G^+_\text{BTZ}(x, x^\prime)=\sum^{+\infty}_{n=-\infty}G^+_\text{AdS}(x, \Gamma^nx^\prime),
\end{eqnarray}
 where $G^+_\text{AdS}(x, x^\prime)$ is the Wightman function in $\mathrm{AdS}_3$ spacetime and $\Gamma\,x^\prime$ represents the action of the identification $\phi\longmapsto\phi+2\pi$ on point $x^\prime$. Assuming that the field satisfies different boundary conditions at spatial infinity due to lack of global 
 hyperbolicity of the BTZ spacetime, the Wightman function can be found analytically as follows \cite{PhysRevD.49.1929}
 \begin{eqnarray}\label{BTZ-WF}
 G^+_\text{BTZ}(x, x^\prime)=\frac{1}{4\sqrt{2}\pi\ell}\sum^\infty_{n=-\infty}\bigg[\frac{1}{\sqrt{\sigma_n}}-\frac{\zeta}{\sqrt{\sigma_n+2}}\bigg],
 \end{eqnarray}
where
\begin{eqnarray}
\sigma_n=\frac{rr^\prime}{r^2_\text{h}}\cosh\bigg[\sqrt{M}(\Delta\phi-2\pi\,n)\bigg]-1-\frac{\sqrt{(r^2-r^2_\text{h})(r^{\prime2}-r^2_\text{h})}}{r^2_\text{h}}\cosh\bigg[\frac{r_\text{h}}{\ell^2}\Delta\,t\bigg],
\end{eqnarray}
with $\Delta\phi=\phi-\phi^\prime$, and $\Delta\,t=t-t^\prime$. The parameter $\zeta\in\{1, 0, -1\}$ respectively specifies the Dirichlet ($\zeta=1$), transparent ($\zeta=0$), and Neumann ($\zeta=-1$) boundary conditions satisfied by the field at spatial infinity \cite{PhysRevD.18.3565}. 
Actually, the Wightman function in the BTZ spacetime is derived in terms of the corresponding Wightman function 
in $\mathrm{AdS}_3$ spacetime as shown in \eqref{Wightman-F}. This is based on the feature of the BTZ solution that the solution arises from identifying points in $\mathrm{AdS}_3$ space, using the orbits of a spacelike Killing vector field \cite{PhysRevD.49.1929}. However, since $\mathrm{AdS}_3$ is not a globally hyperbolic spacetime, 
it means that, in order to have well defined quantum field theory, appropriate boundary conditions must be applied to the field at spatial infinity. Or information can escape or leak in through this time-like surface in a finite coordinate time, spoiling the composition law property of the field propagator. This problem carries over to the black hole solution \cite{PhysRevD.49.1929}. Usually, the simplest boundary conditions are either Dirichlet (where the field vanishes on the boundary) or Neumann (where the normal derivative of the field vanishes on the boundary) \cite{PhysRevD.49.1929, PhysRevD.18.3565}, which means that information is reflected and not lost at the spatial infinity. 
It is also possible to define a quantization scheme on $\mathrm{AdS}_3$ space without using boundary conditions, which is referred to as 
``transparent" boundary conditions" in Ref. \cite{PhysRevD.18.3565}. However, in such case an ``effective Cauchy surface"  for $\mathrm{AdS}$ is needed to be constructed to 
recirculate the energy, angular momentum, etc., lost to timelike infinity, resulting in a well-defined, if rather unusual, conservation law (see more details in Ref. \cite{PhysRevD.18.3565}). Different boundary conditions mean different physical properties of spatial infinity, quantum field at which behaves quite differently due to spacetime scattering, leading to different physical phenomena, e.g., significant effect on the transition rate for Unruh-DeWitt particle detector \cite{PhysRevD.86.064031} and the Fischer information \cite{FIBH}. Therefore, it is of interest to study how different boundary conditions affect charging performance of the QB, and in turn, how to use such effects to explore the properties of spacetime.

In the following discussions, we assume that the QB is spatially fixed at a constant $r$ in the BTZ spacetime such that $\Delta\phi=0$, $\Delta\tau=\tau-\tau^\prime=\sqrt{-g_{00}}\Delta\,t$. In this case, inserting the BTZ Wightman function from \eqref{BTZ-WF} into the QB state parameter \eqref{SP},
we can find
\begin{eqnarray}\label{T-D}
\nonumber
\gamma_+&=&\frac{\mu^2\cos^2(\theta-\Theta)}{2}\bigg(\frac{1}{e^{\Omega/T}+1}\bigg)\sum^{n=\infty}_{n=-\infty}
\bigg[P_{-\frac{1}{2}+i\frac{\Omega}{2\pi\,T}}(\cosh\alpha_n)-\zeta\,P_{-\frac{1}{2}+i\frac{\Omega}{2\pi\,T}}(\cosh\beta_n)\bigg],
\\          \nonumber
\gamma_-&=&\frac{\mu^2\cos^2(\theta-\Theta)}{2}\bigg(\frac{1}{e^{-\Omega/T}+1}\bigg)\sum^{n=\infty}_{n=-\infty}
\bigg[P_{-\frac{1}{2}-i\frac{\Omega}{2\pi\,T}}(\cosh\alpha_n)-\zeta\,P_{-\frac{1}{2}-i\frac{\Omega}{2\pi\,T}}(\cosh\beta_n)\bigg],
\\
\gamma_z&=&\frac{\mu^2\sin^2(\theta-\Theta)}{4}\sum^{n=\infty}_{n=-\infty}
\bigg[P_{-\frac{1}{2}}(\cosh\alpha_n)-\zeta\,P_{-\frac{1}{2}}(\cosh\beta_n)\bigg],
\end{eqnarray}
where $P_\nu(x)$ represents the associated Legendre function of the first kind \cite{2015867}, satisfying $P_{-\frac{1}{2}+i\lambda}=P_{-\frac{1}{2}-i\lambda}$,
$T$ is the local KMS temperature given by
\begin{eqnarray}
T=\frac{r_\text{h}}{2\pi\ell\sqrt{r^2-r^2_\text{h}}},
\end{eqnarray}
and the auxiliary functions $\cosh\alpha_n$ and $\cosh\beta_n$ have been defined as
\begin{eqnarray}
\cosh\alpha_n&=&\frac{r^2_\text{h}}{r^2-r^2_\text{h}}\bigg[\frac{r^2}{r^2_\text{h}}\cosh\big(2\pi\,n\sqrt{M}\big)-1\bigg],
\\
\cosh\beta_n&=&\frac{r^2_\text{h}}{r^2-r^2_\text{h}}\bigg[\frac{r^2}{r^2_\text{h}}\cosh\big(2\pi\,n\sqrt{M}\big)+1\bigg].
\end{eqnarray}
Let us note that the local temperature of the BTZ spacetime (we call this temperature the local Hawking temperature) can be rewritten in the following form
\begin{eqnarray}\label{HT}
T=\frac{\sqrt{a^2_\text{BTZ}-\ell^{-2}}}{2\pi}
\end{eqnarray}
with $a_\text{BTZ}=r/(\ell\sqrt{r^2-r^2_\text{h}})$ representing the acceleration of the constant $r$ trajectory in the BTZ spacetime. Here
$T$ is analogous to the temperature felt by an accelerated observer in $\mathrm{AdS}_3$ spacetime \cite{Jennings_2010}. It is demonstrated in Ref. \cite{Jennings_2010}
that there exists a critical acceleration, $1/\ell$, in $\mathrm{AdS}_3$ spacetime, and only the observer with a constant super-critical acceleration 
$a$ (i.e., $a>1/\ell$) can register the quasi-thermal response with a temperature equal to $\sqrt{a^2-\ell^{-2}}/(2\pi)$. Besides, from the response functions
$\gamma_\pm$ in \eqref{T-D} we can see that the Fermi-Dirac type distribution is displayed as noted in Ref. \cite{PhysRevD.49.1929}.
The response functions in \eqref{T-D} also are the function of various parameters of the QB itself and the BTZ black hole listed in the Table \ref{Table1}. 
\begin{table}[h]
\centering 
\begin{tabular}{| l |c|}
\hline
$\mu$    &   the coupling constant between the QB and field  \\ \hline
$\Delta$    & the QB's energy-level spacing  \\    \hline
$A$    &   the charging amplitude characterizing charing strength of the charger  \\ \hline
$T=\frac{r_\text{h}}{2\pi\ell\sqrt{r^2-r^2_\text{h}}}$     & the local KMS temperature of BTZ black hole   \\    \hline
$r_\text{h}=\sqrt{M}\ell$   &  the horizon radius with $M$ being the mass of BTZ black hole  \\    \hline 
$\Lambda=-1/\ell^2$   &  the negative cosmological constant  \\    \hline 
$r$   &  the distance of the observer (or QB) away from the BTZ black hole \\     \hline 
$\zeta$   &  the boundary condition parameter  \\     \hline 
\end{tabular}
\caption{Various parameters of both the QB and BTZ black hole. All of them will affect the dynamics of the QB and thus the charging performance.}
\label{Table1}
\end{table}
Clearly, these parameters determine the dissipation and thus affect the dynamics of the QB. In what follows, how the different parameters affect charging performance of the QB will be studied.

\subsection{Charging dynamics}

In this section we will study the charging performances in the presence of dissipation which results from the vacuum fluctuations in the BTZ spacetime.

We now use the previous results in order to describe the dynamics of the QB by setting different initial parameters. We will determine the energy variation associated to 
the charging process by considering the QB is initially prepared in the ground state $|g\rangle$, i.e., $w_1(0)=0, w_2(0)=0, w_3(0)=-1$ in  $\rho_\text{QB}(0)$.
Unless otherwise stated, we will discuss the two limiting case of decoherence coupling ($\theta=0$ in Eq. \eqref{Interaction-H}) and of pure dephasing coupling ($\theta=\pi/2$ in Eq. \eqref{Interaction-H}). 

Before discussing the dissipative dynamics, let us recall the scenario where the dissipation is absent, which corresponds to $\mu=0$ case in the interaction Hamiltonian \eqref{Interaction-H}. In this case, all energy supplied by $H_\text{C}$ is transferred to the QB, whose maximum energy that can be stored is given by the energy level spacing $\Delta$. The QB then is charged by a static driving protocol according to \cite{Binder_2015, Crescente_2020}
\begin{eqnarray}\label{EQB0}
\langle\,E_\text{QB}(\tau)\rangle=\frac{\Delta}{2}\frac{A^2}{\Omega^2}[1-\cos(\Omega\tau)],
\end{eqnarray}
where $\Omega=\sqrt{\Delta^2+A^2}$ is the Rabi frequency. From \eqref{EQB0} we can see that since the evolution of the closed QB is unitary, the charging and discharge processes of the  QB behaves  periodically. It is completely discharged (again empty and in the $|g\rangle$ state) when $\Omega\tau=2\pi\,n$ ($n>0$ integer). Note that the optimal charging time, when the energy stored reaches its maxima, is $\Omega\tau=\pi\,n$ ($n>0$ odd numbers). In this case, the energy stored 
in the QB is 
\begin{eqnarray}\label{MQBE}
\langle\,E_\text{QB}(\tau)\rangle_\text{max}=\Delta\frac{A^2}{\Omega^2}.
\end{eqnarray}
This amount approaches its maxima for very large amplitude $A\gg\Delta$.

In the presence of dissipation, we first consider the case of decoherence coupling ($\theta=0$ in Eq. \eqref{Interaction-H}), where the QB-quantum field coupling only has 
the $\sigma_x$ component. We find the corresponding average energy associated to the QB is given by
\begin{eqnarray}\label{AEBD}
\nonumber
\langle\,E_\text{QB}(\tau)\rangle&=&\frac{\Delta}{2}\bigg\{1-\frac{\Delta^2}{\Omega^2}e^{-\Gamma^{\text{r}}_0\tau}-
\frac{\Delta}{\Omega}\bigg(\frac{e^{\Omega/T}-1}{e^{\Omega/T}+1}\bigg)\big(1-e^{-\Gamma^{\text{r}}_0\tau}\big)-\frac{A^2}{\Omega^2}\cos(\Omega\tau)
e^{-\Gamma_0\tau}\bigg\},
\\
\end{eqnarray}
where 
\begin{eqnarray}\label{rates1}
\nonumber
\Gamma^{\text{r}}_0&=&\frac{\mu^2\Delta^2}{2\Omega^2}\sum^{n=\infty}_{n=-\infty}
\bigg[P_{-\frac{1}{2}+i\frac{\Omega}{2\pi\,T}}(\cosh\alpha_n)-\zeta\,P_{-\frac{1}{2}+i\frac{\Omega}{2\pi\,T}}(\cosh\beta_n)\bigg],
\\
\Gamma_0&=&\frac{1}{2}\Gamma^{\text{r}}_0+\frac{\mu^2A^2}{2\Omega^2}\sum^{n=\infty}_{n=-\infty}
\bigg[P_{-\frac{1}{2}}(\cosh\alpha_n)-\zeta\,P_{-\frac{1}{2}}(\cosh\beta_n)\bigg].
\end{eqnarray}
Similarly, the energy variation of the charger $\langle\,E_\text{C}(\tau)\rangle$ reads 
\begin{eqnarray}\label{AECD}
\langle\,E_\text{C}(\tau)\rangle&=&\frac{A^2}{2\Omega}\bigg\{-\frac{\Delta}{\Omega}e^{-\Gamma^{\text{r}}_0\tau}-
\bigg(\frac{e^{\Omega/T}-1}{e^{\Omega/T}+1}\bigg)\big(1-e^{-\Gamma^{\text{r}}_0\tau}\big)+\frac{\Delta}{\Omega}\cos(\Omega\tau)
e^{-\Gamma_0\tau}\bigg\}.
\end{eqnarray}
Note that dissipation effects are reflected in the incoherent relaxation rate, $\Gamma^{\text{r}}_0$, and dephasing rate, $\Gamma_0$.

For the pure dephasing coupling case, it is described by setting $\theta=\pi/2$ in Eq. \eqref{Interaction-H}, where the QB-quantum field coupling only has the $\sigma_z$ component. This coupling case may lead to a different dissipative dynamics compared with the decoherence case. 
Indeed, the corresponding average energy associated to the QB in this case can be found as
\begin{eqnarray}\label{AEBP}
\nonumber
\langle\,E_\text{QB}(\tau)\rangle&=&\frac{\Delta}{2}\bigg\{1-\frac{\Delta^2}{\Omega^2}e^{-\Gamma^{\text{r}}_{\pi/2}\tau}-
\frac{\Delta}{\Omega}\bigg(\frac{e^{\Omega/T}-1}{e^{\Omega/T}+1}\bigg)\big(1-e^{-\Gamma^{\text{r}}_{\pi/2}\tau}\big)-\frac{A^2}{\Omega^2}\cos(\Omega\tau)
e^{-\Gamma_{\pi/2}\tau}\bigg\},
\\
\end{eqnarray}
and the charger contribution is given by
\begin{eqnarray}\label{AECP}
\langle\,E_\text{C}(\tau)\rangle&=&\frac{A^2}{2\Omega}\bigg\{-\frac{\Delta}{\Omega}e^{-\Gamma^{\text{r}}_{\pi/2}\tau}-
\bigg(\frac{e^{\Omega/T}-1}{e^{\Omega/T}+1}\bigg)\big(1-e^{-\Gamma^{\text{r}}_{\pi/2}\tau}\big)+\frac{\Delta}{\Omega}\cos(\Omega\tau)
e^{-\Gamma_{\pi/2}\tau}\bigg\},
\end{eqnarray}
where 
\begin{eqnarray}\label{DR}
\nonumber
\Gamma^{\text{r}}_{\pi/2}&=&\frac{\mu^2A^2}{2\Omega^2}\sum^{n=\infty}_{n=-\infty}
\bigg[P_{-\frac{1}{2}+i\frac{\Omega}{2\pi\,T}}(\cosh\alpha_n)-\zeta\,P_{-\frac{1}{2}+i\frac{\Omega}{2\pi\,T}}(\cosh\beta_n)\bigg],
\\
\Gamma_{\pi/2}&=&\frac{1}{2}\Gamma^{\text{r}}_{\pi/2}+\frac{\mu^2\Delta^2}{2\Omega^2}\sum^{n=\infty}_{n=-\infty}
\bigg[P_{-\frac{1}{2}}(\cosh\alpha_n)-\zeta\,P_{-\frac{1}{2}}(\cosh\beta_n)\bigg].
\end{eqnarray}

Let us turn our attention to analytically evaluating the average energy in some special cases. In the infinite time limit, i.e., $\tau\rightarrow\infty$, for both the
decoherence and pure dephasing coupling cases the average energies in \eqref{AEBD}, \eqref{AECD}, \eqref{AEBP}, \eqref{AECP} approach to 
\begin{eqnarray}\label{QBE}
\nonumber
\langle\,E_\text{QB}(\infty)\rangle&=&\frac{\Delta}{2}\bigg\{1-\frac{\Delta}{\Omega}\bigg(\frac{e^{\Omega/T}-1}{e^{\Omega/T}+1}\bigg)\bigg\},
\\
\langle\,E_\text{C}(\infty)\rangle&=&-\frac{A^2}{2\Omega}\bigg(\frac{e^{\Omega/T}-1}{e^{\Omega/T}+1}\bigg).
\end{eqnarray}
We note that in the infinite evolution time limit the average energies for the QB and charger are independent of the kinds of dissipation. Besides, in this case the
average energies do not depend on the charging time, and is quite different from the case without dissipation shown in \eqref{EQB0}, which is alway periodic with
the charging time. 

\begin{figure*}[t]
\centering
\includegraphics[width=0.9  \textwidth]{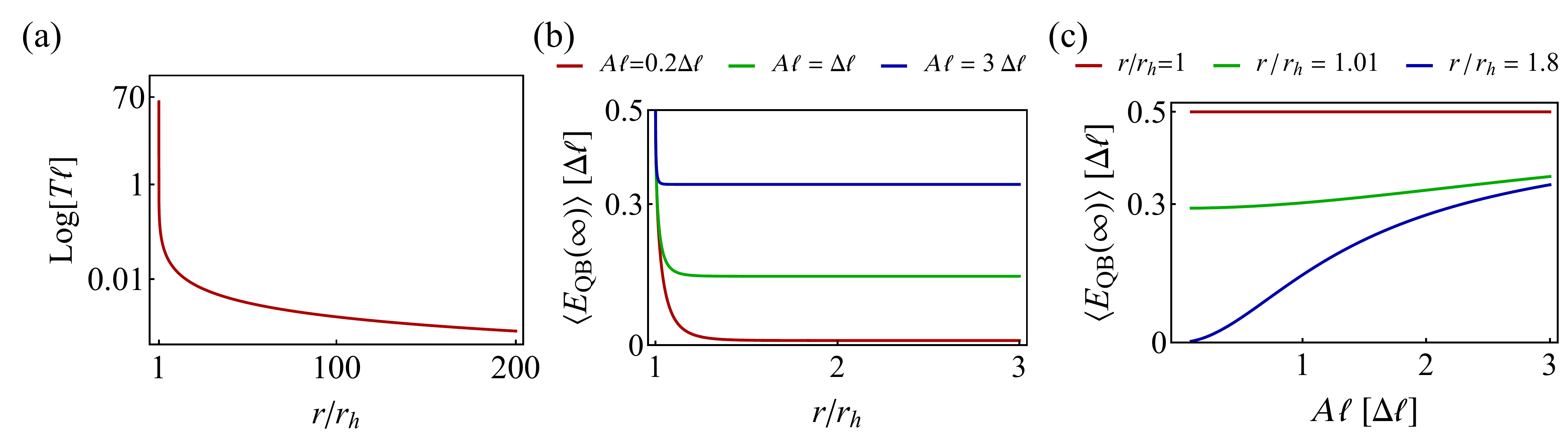}
\caption{(a) The local KMS temperature of BTZ black hole shown in Eq. \eqref{HT} as a function of the relative distance $r/r_h$ of the QB; (b) The average energy stored in the QB
$\langle\,E_\text{QB}(\infty)\rangle$ shown in Eq. \eqref{QBE} (in units $\Delta\ell$) as a function of the relative distance $r/r_h$; (c) The average energy stored in the QB $\langle\,E_\text{QB}(\infty)\rangle$ shown in Eq. \eqref{QBE} (in units $\Delta\ell$) as a function of the charging amplitude $A\ell$.}\label{fig1}
\end{figure*}

In Fig. \ref{fig1} we show the local KMS temperature of BTZ black hole \eqref{HT} and the average energy \eqref{QBE} stored in the QB in the infinite charging time. 
The temperature decreases monotonously as the increase of the distance of the QB away from the BTZ black hole horizon $r/r_\text{h}$. It means that the further the QB is located away from the black horizon, the lower the temperature that the QB feels is. In the open quantum system theory, the thermal bath with higher temperature will lead to stronger dissipation of the QB (see more details below). In the infinite charging time limit, the average energy shored in the QB approaches to $\lim_{r/r_h\rightarrow1}\langle\,E_\text{QB}(\infty)\rangle=\frac{\Delta}{2}$ when the QB is located near the horizon, which does not depend on the charging amplitude $A$. This suggests that when the temperature of thermal bath (or local Hawking temperature) is extremely high the stationary average energy that is deposited in the QB through charging only depends on its energy-level spacing, while the other information of external environment, such as the charging amplitude and black hole background parameters, is not encoded in the QB in this scenario. Besides, this average energy decreases sharply with the increase of the relative distance $r/r_\text{h}$ when near the horizon. That is to say, in the extremely high-temperature thermal bath, even small temperature fluctuation would lead to strong dissipation and thus the drastic change of the average energy stored in the QB. While when the QB is far way from the horizon (the local Hawking temperature that the QB feels approaches to zero), the average energy approaches to a constant $\lim_{r/r_\text{h}\rightarrow\infty}\langle E_\text{QB}(\infty)\rangle=\frac{\Delta}{2}(1-\frac{\Delta}{\Omega})$. It depends on the charging amplitude and is different from the case near the horizon. We can find that when QB far away from the horizon the stronger the charging amplitude is, the more the average energy stored in the QB is. This means that, in the presence of dissipation from vacuum fluctuations of external environment, larger driving amplitude $A$ leads to batter charging of the QB.
We can also see this behavior in Fig. \ref{fig1} (c). Furthermore, 
the further the QB locates away from the black hole horizon, the more strongly its average energy in the infinite time limit depends on the charging amplitude $A$.
This is because that when far way from the black hole horizon, the local Hawking temperature in \eqref{HT} approaches to vanishment, the dissipation of the QB effectively coming from the vacuum fluctuations of the quantum field could be considered to be weak. In this case the charger thus would play a dominated role in the whole charging process.

\begin{figure*}[t]
\centering
\includegraphics[width=0.9  \textwidth]{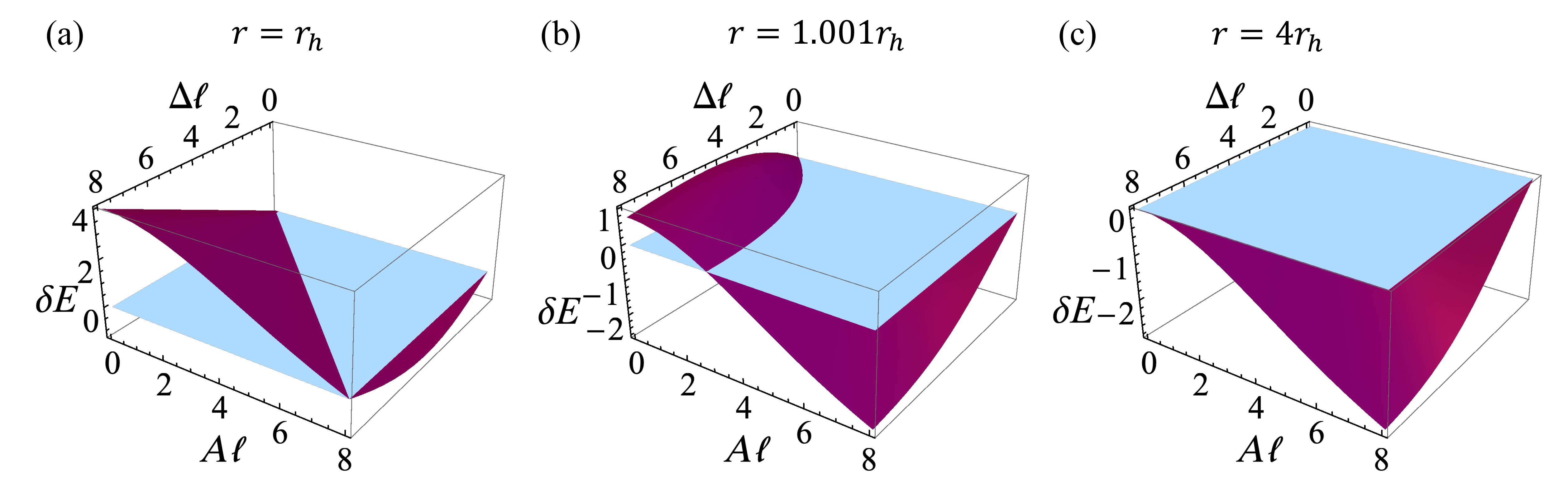}
\caption{The average-energy difference stored in the QB between the dissipation case and the closed QB case shown in \eqref{difference}, as a function of the QB energy-level spacing $\Delta\ell$ and the charging amplitude $A\ell$. The location of the QB is: (a) $r=r_h$; (b) $r=1.001r_h$; (c) $r=4r_h$. The purple surface denotes the average-energy difference, while the bluish surface denotes $\delta\,E=0$ surface.}\label{fig2}
\end{figure*}

An interesting question is that in the presence of dissipation from vacuum field fluctuations in the BTZ spacetime, whether 
the maximum average energy stored in QB could be enhanced. We will compare $\langle\,E_\text{QB}(\infty)\rangle$ shown in \eqref{QBE} with $\langle\,E_\text{QB}(\tau)\rangle_\text{max}$ shown in \eqref{MQBE}, through defining 
\begin{eqnarray}\label{difference}
\delta E=\langle\,E_\text{QB}(\infty)\rangle-\langle\,E_\text{QB}(\tau)\rangle_\text{max}=\frac{\Delta}{2}\bigg[1-\frac{\Delta}{\Omega}\bigg(\frac{e^{\Omega/T}-1}{e^{\Omega/T}+1}\bigg)-\frac{2A^2}{\Omega^2}\bigg].
\end{eqnarray}
When $A\ge\Delta$, we can always find that the average-energy difference in this case is smaller than zero, $\delta E\le0$. Which means that the dissipation from vacuum field fluctuations in the BTZ spacetime might causes a worse charging performance compared with the closed QB case. Besides, if $T\gg\Omega$, we can find in Eq. \eqref{difference} $\delta E\simeq\frac{\Delta}{2}(1-2A^2/\Omega^2)=\Delta(\Delta^2-A^2)/2\Omega^2$. While for $T\ll\Omega$, we can find $\delta E\simeq\frac{\Delta}{2}(1-\Delta/\Omega-2A^2/\Omega^2)$. In these two cases, whether the charging 
performance in the presence of dissipation is better or worse than the closed QB charging case depends on the QB energy-level spacing $\Delta$ and the charging amplitude $A$. In Fig. \ref{fig2} we plot the average-energy difference shown in \eqref{difference} as a function of the QB energy-level spacing $\Delta\ell$ and the charging amplitude $A\ell$. We can find that the location of the QB (or equivalently the local Hawking temperature), $r$ (or $T$), affects the charging performance very much. When the QB is relatively far away from the horizon $r_h$, the average energy stored in QB in the presence of dissipation would never be more than 
that for the closed QB case. Thus when the local Hawking temperature is relatively smaller, the dissipation from vacuum field fluctuations in the BTZ spacetime causes harmful influence on the charging performance. However, we can also find that when the QB approaches to the horizon, where the local Hawking temperature is higher,
the average energy shored in the QB in the presence of dissipation might be more than that for the closed QB case under the condition, $T\ge\Omega/\ln\big[\frac{\Delta\Omega+\Delta^2-A^2}{\Delta\Omega+A^2-\Delta^2}\big]$ with $\Delta>A$. It means in this case the dissipation from vacuum field fluctuations in the BTZ spacetime might enhance the charging performance. In other words, one can extract the energy from the vacuum fluctuations in curved space through dissipation in the charging process.

\subsection{Effect of dissipation on average energy} 

We now discuss the results obtained above, showing how the effects of dissipation from vacuum field fluctuations in the BTZ spacetime modify the charging dynamics of the QB. 

\subsubsection{$\Delta<A$ case}

\begin{figure*}[t]
\centering
\includegraphics[width=0.9  \textwidth]{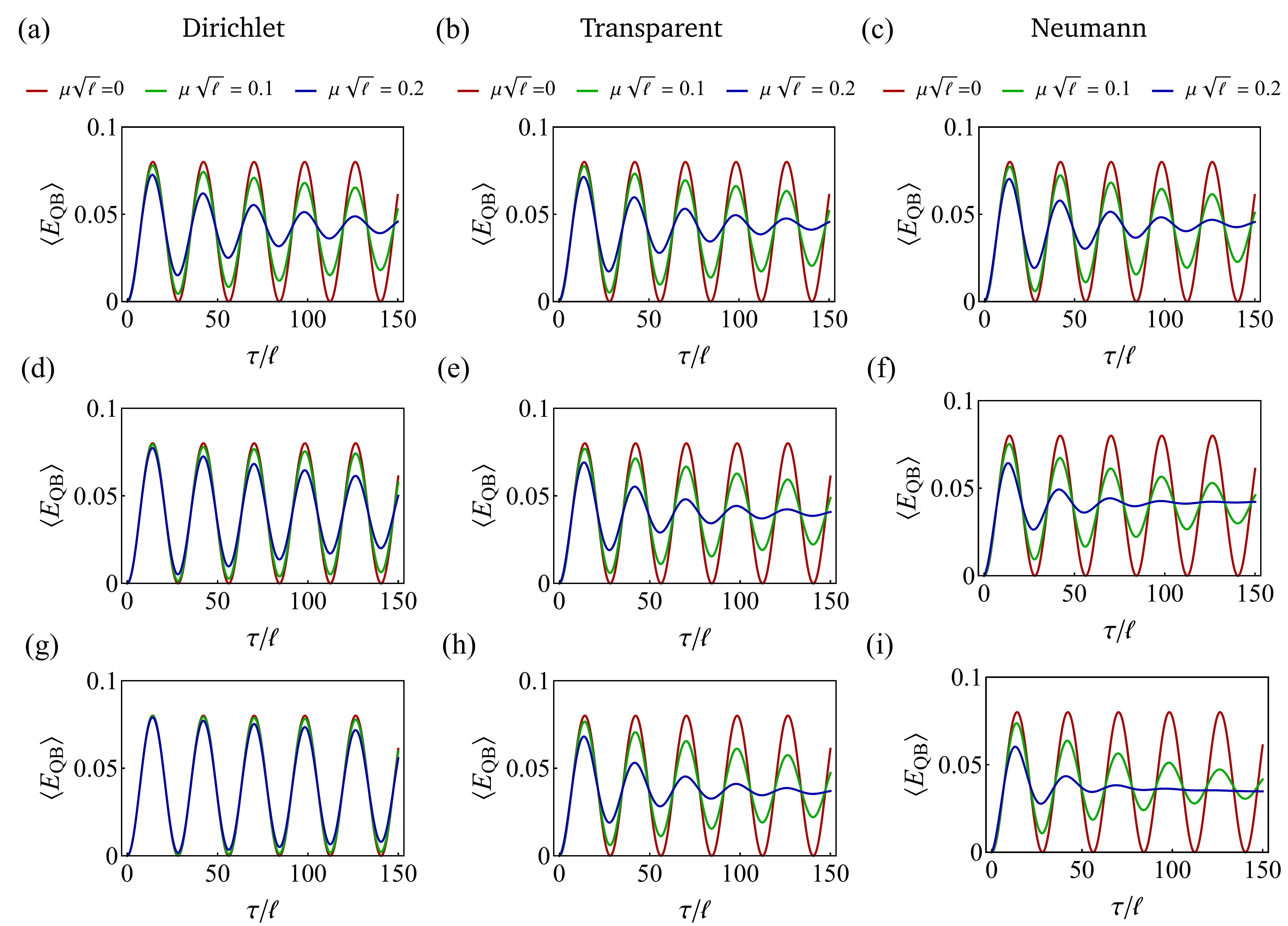}
\caption{Time evolution of the average energy stored in the QB $\langle\,E_\text{QB}(\tau)\rangle$ shown in Eq. \eqref{AEBD} for the decoherence coupling case with $\theta=0$. Three different boundary conditions in BTZ spacetime are shown for various coupling strength $\mu\sqrt{\ell}$. Here we have taken
the energy-level spacing $\Delta\ell=0.1$, the charging amplitude $A\ell=0.2$, and the mass $M=1$. The first, second, and third row panels have taken 
the relative distance of the QB in BTZ spacetime $r/r_h=1.001$, $r/r_h=1.2$ and $r/r_h=1.8$, respectively.}\label{fig3}
\end{figure*}

In Fig. \ref{fig3} we show the time evolution of the average energy stored in the QB $\langle\,E_\text{QB}(\tau)\rangle$ shown in Eq. \eqref{AEBD} for the decoherence coupling case 
($\theta=0$ case). We fixed the energy-level spacing of the QB $\Delta\ell=0.1$, the charging amplitude $A\ell=0.2$, and the BTZ black hole mass $M=1$. Representative examples
of different distance of the QB (or the local Hawking temperature) away from the horizon $r/r_h=1.001$ (Fig. \ref{fig3} (a), (b) and (c)), $r/r_h=1.2$ (Fig. \ref{fig3} (d), (e) and (f)), and $r/r_h=1.8$ (Fig. \ref{fig3} (g), (h) and (i)) have been chosen. The closed QB system driven by a static bias (see Eq. \eqref{EQB0}) is also reported as a reference limit (see red curves). It is found that, in the presence of dissipation, the charged average energy presents a damped oscillatory behavior, whose amplitude is modulated by the exponential decay dictated by the incoherent relaxation and dephasing rates given in Eqs. \eqref{rates1}. With the increase of the local Hawking temperature, the damping strength is enhanced, and the average energy stored in the QB degrades faster to its asymptotic limit \eqref{QBE} for the Dirichlet boundary case, while it is opposite for the transparent and the Neumann boundary cases. 
The coupling strength 
$\mu\sqrt{\ell}$ could also enhance the dissipation as expected, showing stronger damping with the increase of the coupling strength. Besides, boundary conditions play an important role in the charging process. Although the boundary conditions do not change the average energy in the infinite charging time limit shown in 
\eqref{QBE}, they could lead to different damping. We can see that the Neumann boundary condition causes stronger damping than the transparent one, while the transparent one induces stronger damping than the Dirichlet one. This difference is more clear when the local Hawking temperature becomes smaller.

\begin{figure*}[t]
\centering
\includegraphics[width=0.9  \textwidth]{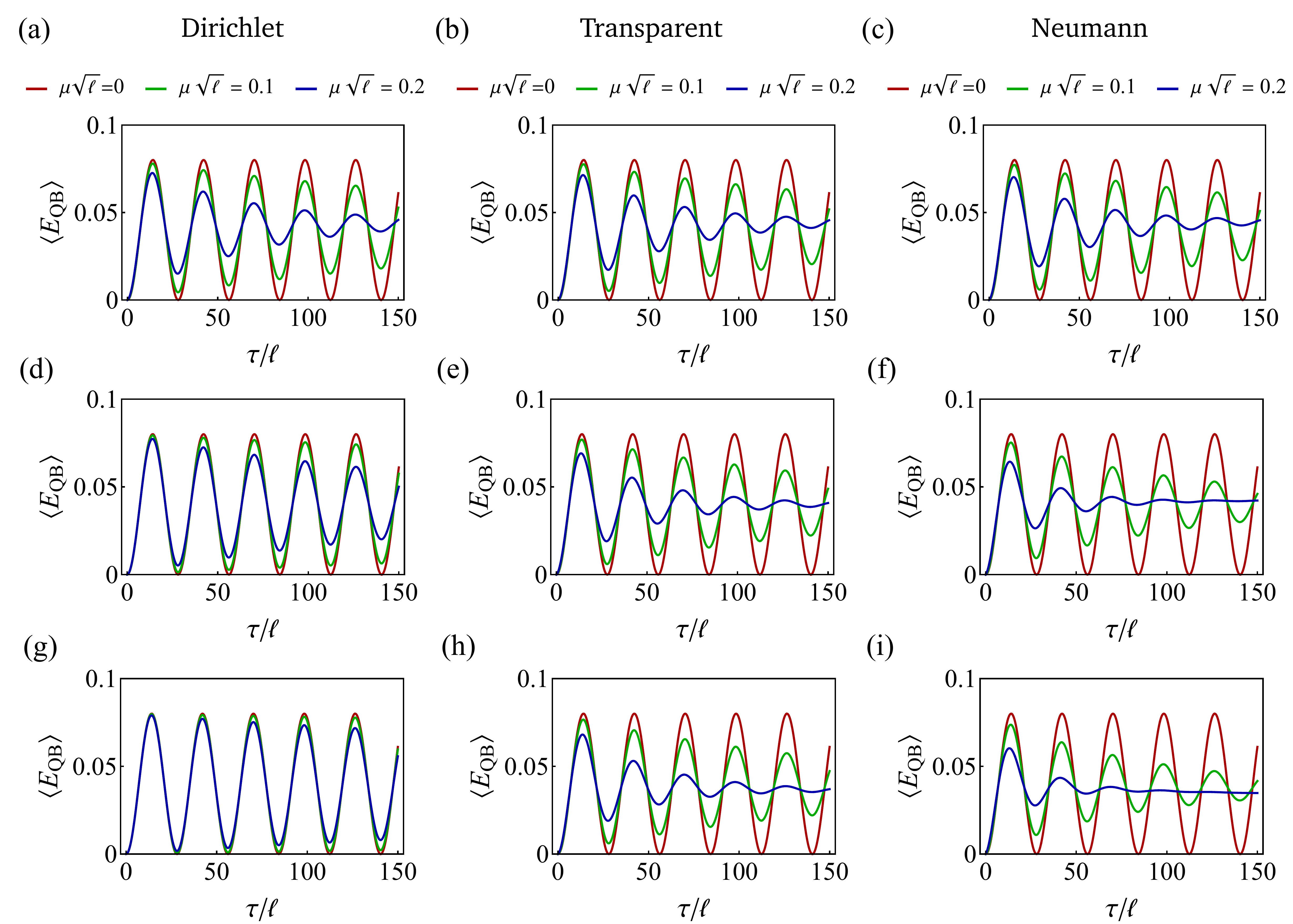}
\caption{Time evolution of the average energy stored in the QB $\langle\,E_\text{QB}(\tau)\rangle$ shown in Eq. \eqref{AEBP} for the pure dephasing coupling case with $\theta=\pi/2$. Three different boundary conditions in BTZ spacetime are shown for various coupling strength $\mu\sqrt{\ell}$. Here we have taken
the energy level spacing $\Delta\ell=0.1$, the charging amplitude $A\ell=0.2$, and the mass $M=1$. The first, second, and third row panels have taken 
the relative distance of the QB in BTZ spacetime $r/r_h=1.001$, $r/r_h=1.2$ and $r/r_h=1.8$, respectively.}\label{fig4}
\end{figure*}

In Fig. \ref{fig4} we show the time evolution of the average energy stored in the QB $\langle\,E_\text{QB}(\tau)\rangle$ shown in Eq. \eqref{AEBP} for the pure dephasing coupling case with $\theta=\pi/2$. As comparison, we fixed the same energy-level spacing of the QB $\Delta\ell=0.1$, the charging amplitude $A\ell=0.2$, and the BTZ black hole mass $M=1$ as that in Fig. \ref{fig3}. We find that in the presence of dissipation all curves present a damped oscillatory behaviour, whose amplitude is modulated by the exponential decay dictated by the incoherent relaxation and dephasing rates given in Eqs. \eqref{DR}. These rate expressions are different from that for the decoherence coupling case discussed above, and thus may gives rise to different relaxation dynamics as shown in the figure. It is evident that the average energy stored in the QB $\langle\,E_\text{QB}(\tau)\rangle$ shown in Eq. \eqref{AEBD}, when varying boundary conditions, coupling strength, and the relative distance of the QB, shares the same behaviour as that for the decoherence coupling case discussed above. However, they are different quantitatively. 

\begin{figure*}[t]
\centering
\includegraphics[width=0.9  \textwidth]{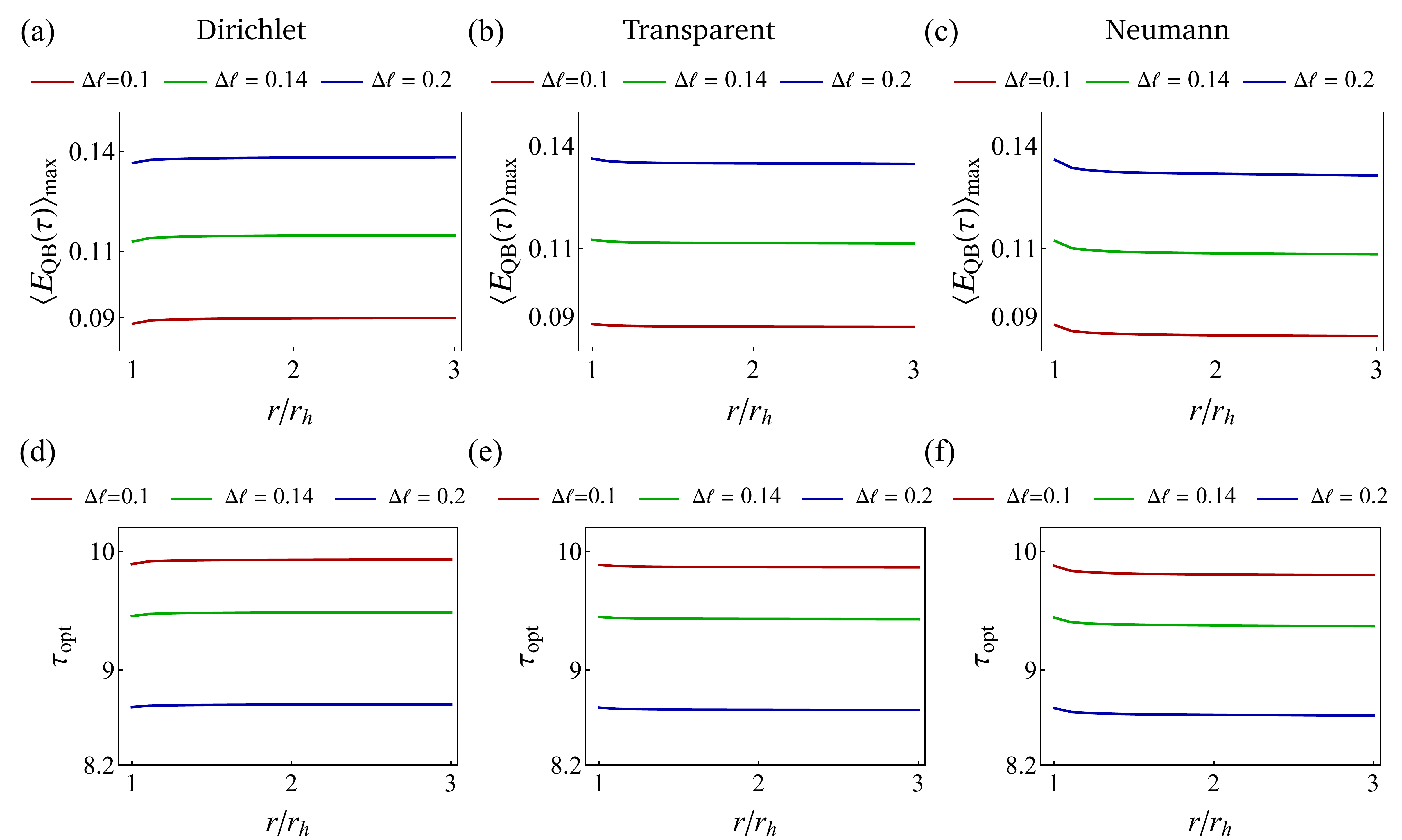}
\caption{The maximum average energy stored in the QB $\langle\,E_\text{QB}(\tau)\rangle_\text{max}$, (a), (b), (c), and the optimal charging time $\tau_\text{opt}$, (d), (e), (f), as a function of the relative distance of the QB in BTZ spacetime $r/r_h$ for the decoherence coupling case ($\theta=0$). Three different boundary conditions in BTZ spacetime are shown for various energy-level spacing $\Delta\ell$. Here we have taken
the coupling strength $\mu\sqrt{\ell}=0.1$, the charging amplitude $A\ell=0.3$, and the mass $M=1$.}\label{fig5}
\end{figure*}

\begin{figure*}[t]
\centering
\includegraphics[width=0.9  \textwidth]{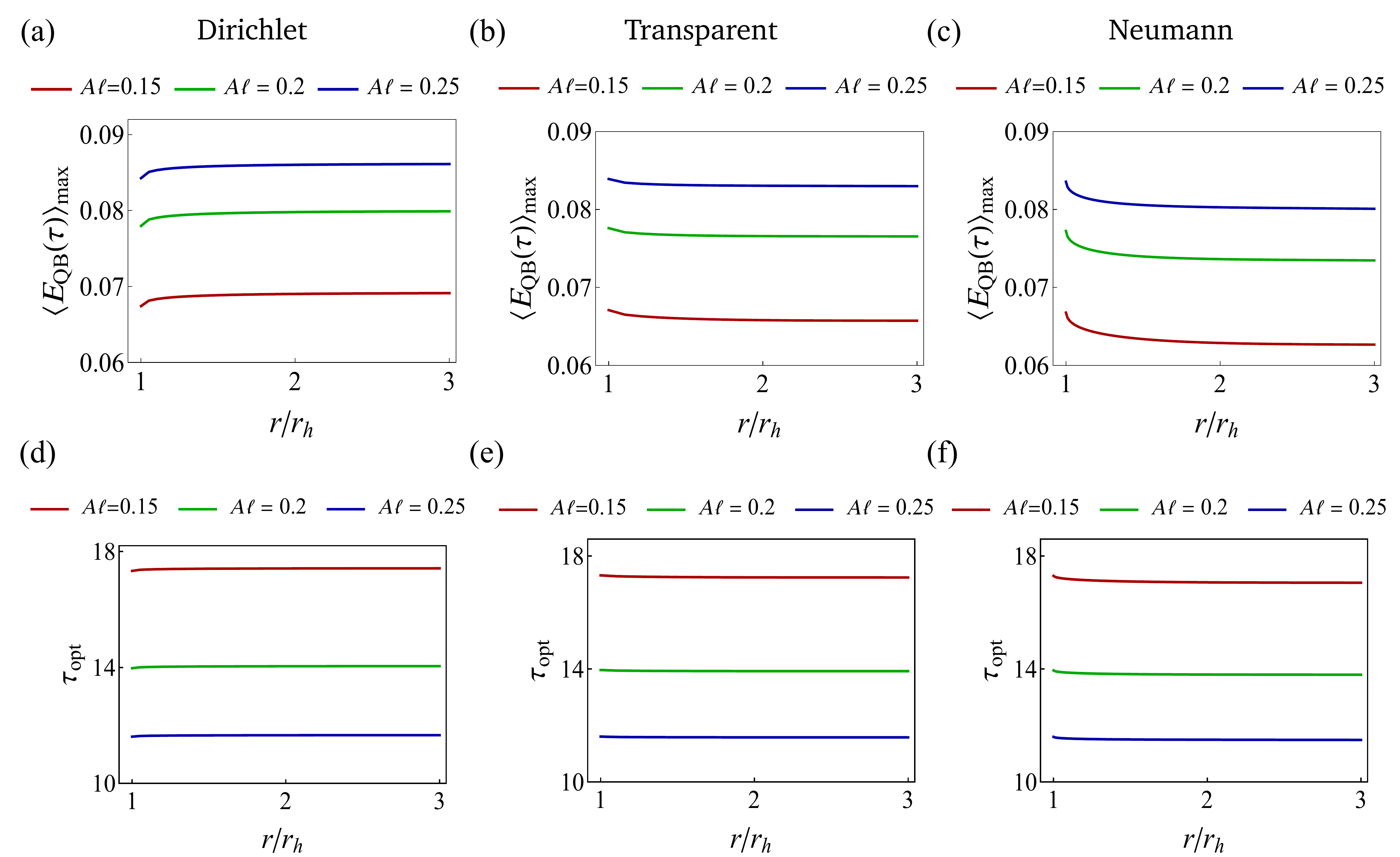}
\caption{The maximum average energy stored in the QB $\langle\,E_\text{QB}(\tau)\rangle_\text{max}$, (a), (b), (c), and the optimal charging time $\tau_\text{opt}$, (d), (e), (f), as a function of the relative distance of the QB in BTZ spacetime $r/r_h$ for the decoherence coupling case ($\theta=0$). Three different boundary conditions in BTZ spacetime are shown for various charging amplitude $A\ell$. Here we have taken
the coupling strength $\mu\sqrt{\ell}=0.1$, the energy-level spacing $\Delta\ell=0.1$, and the mass $M=1$.}\label{fig6}
\end{figure*}

For both the decoherence and the pure dephasing coupling cases, there is an optimal charging time, $\tau_\text{opt}$, at which the average energy stored in the QB takes the maxima. In Figs. \ref{fig5} and \ref{fig6} we plot this maximum average energy and the optimal charging time as a function of the local Hawking temperature parameter
$r/r_h$ by varying the energy-level spacing $\Delta\ell$ and the charging amplitude $A\ell$ for decoherence coupling case ($\theta=0$). We can see from Fig. \ref{fig5} that the boundary condition influences the behavior of the maximum average energy and the optimal charging time vs the local Hawking temperature parameter significantly. For the Dirichlet case, the maximum average energy and the optimal charging time increase with the decrease of the local Hawking temperature (local Hawking temperature decreases with the increase of $r/r_h$, see Fig. \ref{fig1}). When the local Hawking temperature approaches to zero, these two quantities approach to asymptotical value. However, we can find that for both the transparent case and the Neumann case, the maximum average energy and the optimal charging time decrease when the local Hawking temperature decreases. These two quantities, when the local Hawking temperature goes to zero, also approach to asymptotical value. Besides, no matter for which boundary conditions, we can find that when the charging amplitude $A\ell$ and the local Hawking temperature are fixed,
the maximum average energy increases with the increase of the energy-level spacing of the QB $\Delta\ell$, while the corresponding optimal charging time 
decreases. In Fig. \ref{fig6} we fix the energy-level spacing of the QB, $\Delta\ell$, and plot the maximum average energy stored in the QB $\langle\,E_\text{QB}(\tau)\rangle_\text{max}$ and the optimal charging time $\tau_\text{opt}$ as a function of the relative distance of the QB in BTZ spacetime $r/r_h$ for various charging amplitude $A\ell$. We can also find that the boundary condition influences 
the behavior of the maximum average energy and the optimal charging time vs the local Hawking temperature parameter significantly. 
When the relative distance of the QB in BTZ spacetime $r/r_h$ approaches to the infinity, both the maximum average energy and the optimal charging time
approach to asymptotical value. Furthermore, no matter for which boundary conditions, the maximum average energy increases with the increase of the charging amplitude, while the optimal charging time decreases 
when the energy-level spacing of the QB and the local Hawking temperature are fixed.

\begin{figure*}[t]
\centering
\includegraphics[width=0.9  \textwidth]{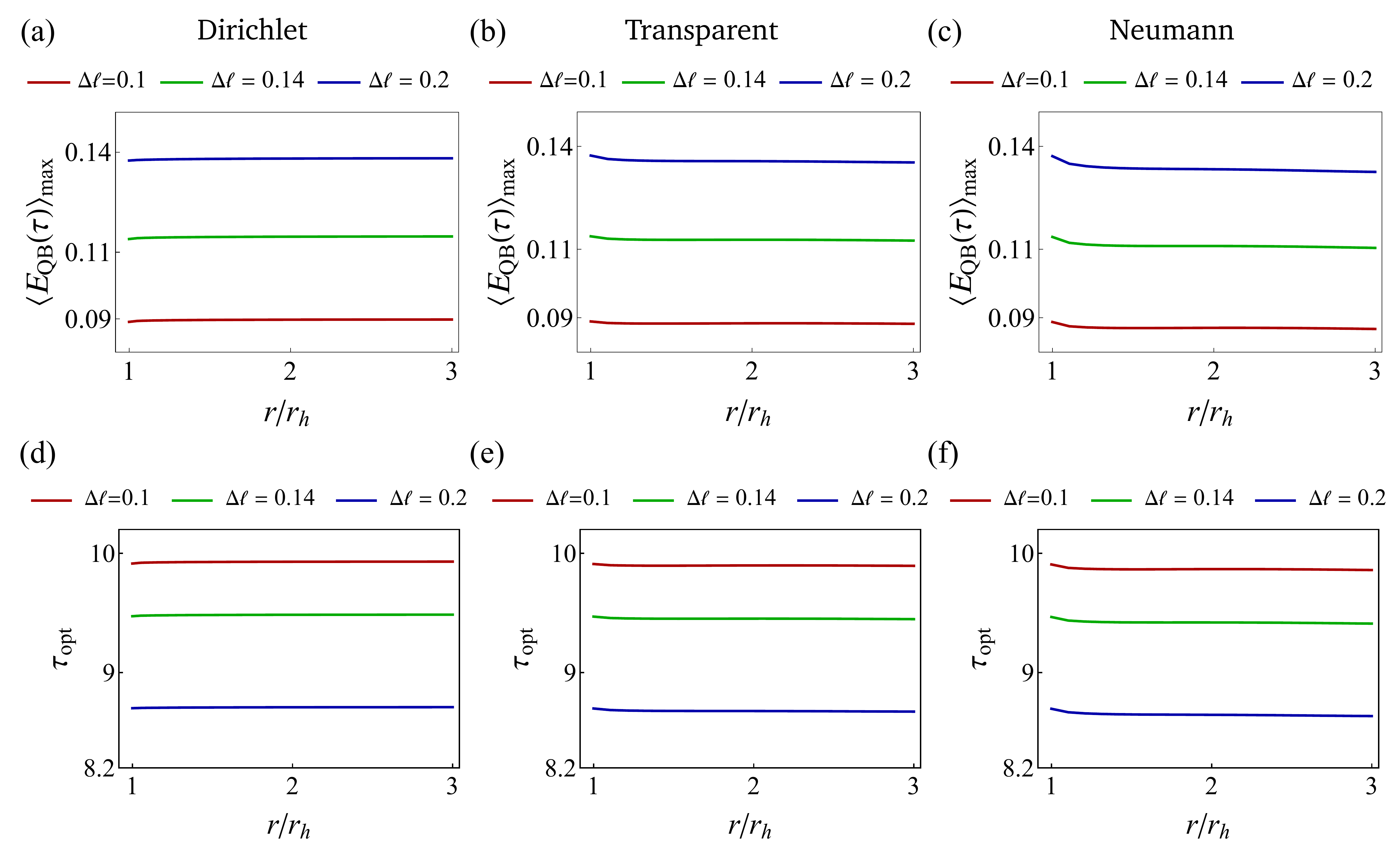}
\caption{The maximum average energy stored in the QB $\langle\,E_\text{QB}(\tau)\rangle_\text{max}$, (a), (b), (c), and the optimal charging time $\tau_\text{opt}$, (d), (e), (f), as a function of the relative distance of the QB in BTZ spacetime $r/r_h$ for pure dephasing coupling case with $\theta=\pi/2$. Three different boundary conditions in BTZ spacetime are shown for various energy-level spacing $\Delta\ell$. Here we have taken
the coupling strength $\mu\sqrt{\ell}=0.1$, the charging amplitude $A\ell=0.3$, and the mass $M=1$.}\label{fig7}
\end{figure*}

\begin{figure*}[t]
\centering
\includegraphics[width=0.9  \textwidth]{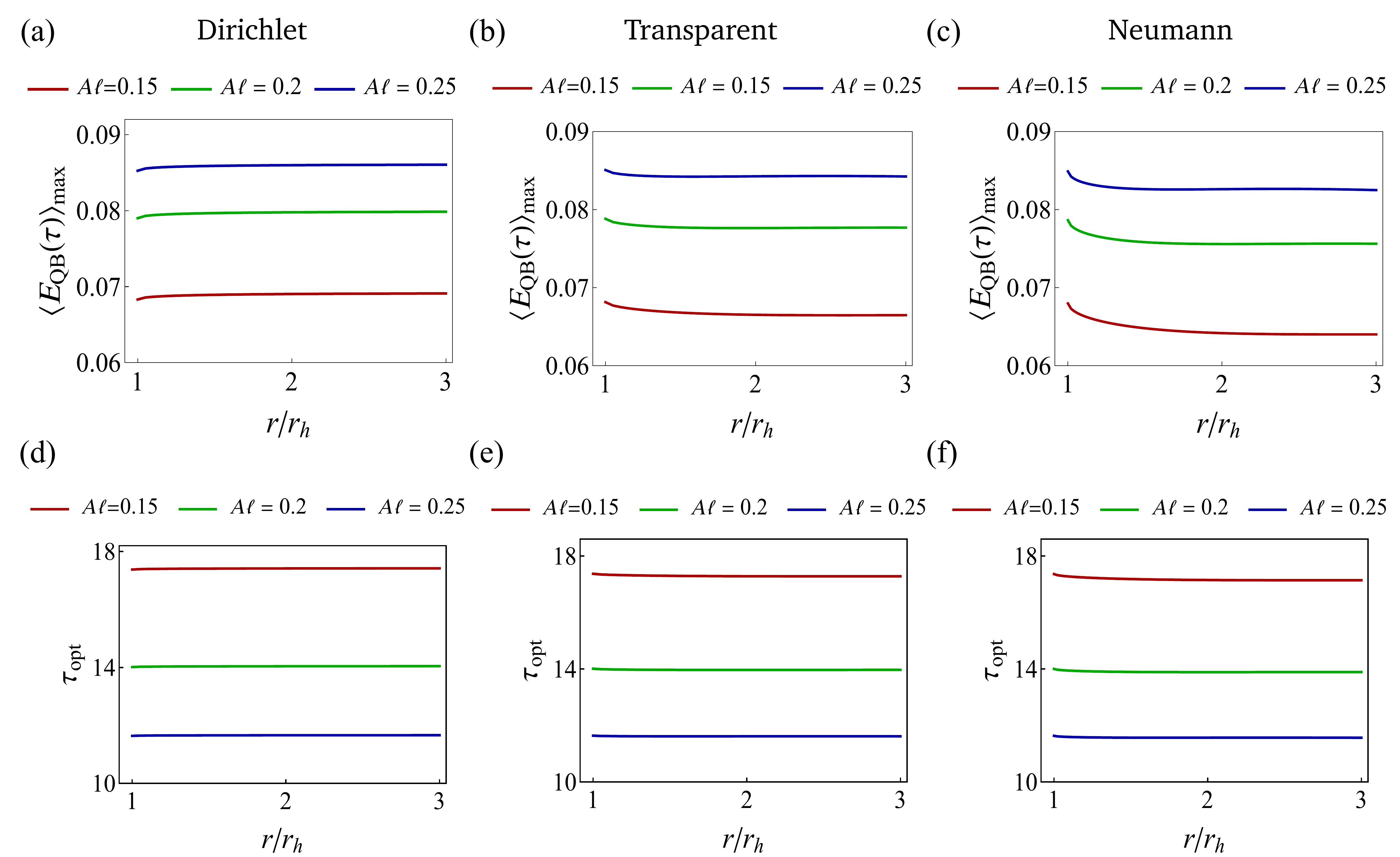}
\caption{The maximum average energy stored in the QB $\langle\,E_\text{QB}(\tau)\rangle_\text{max}$, (a), (b), (c), and the optimal charging time $\tau_\text{opt}$, (d), (e), (f), as a function of the relative distance of the QB in BTZ spacetime $r/r_h$ for pure dephasing coupling case with $\theta=\pi/2$. Three different boundary conditions in BTZ spacetime are shown for various charging amplitude $A\ell$. Here we have taken
the coupling strength $\mu\sqrt{\ell}=0.1$, the energy-level spacing $\Delta\ell=0.1$, and the mass $M=1$.}\label{fig8}
\end{figure*}

In Figs. \ref{fig7} and \ref{fig8} we plot the maximum average energy and the optimal charging time as a function of the local Hawking temperature parameter
$r/r_h$ by varying the energy-level spacing $\Delta\ell$ and the charging amplitude $A\ell$ for pure dephasing coupling case ($\theta=\pi/2$). We can find for this 
pure dephasing coupling case the maximum average energy and the optimal charging time share the same behaviors as that for the decoherence case shown in Figs. \ref{fig5} and \ref{fig6}, while their magnitudes are different. This is because for both the decoherence case and the pure dephasing case their corresponding average energy associated to the QB share the same formula shown in \eqref{AEBD} and \eqref{AEBP}, but with different incoherent relaxation and dephasing rates shown in
\eqref{rates1} and \eqref{DR}.

\begin{figure*}[t]
\centering
\includegraphics[width=0.9 \textwidth]{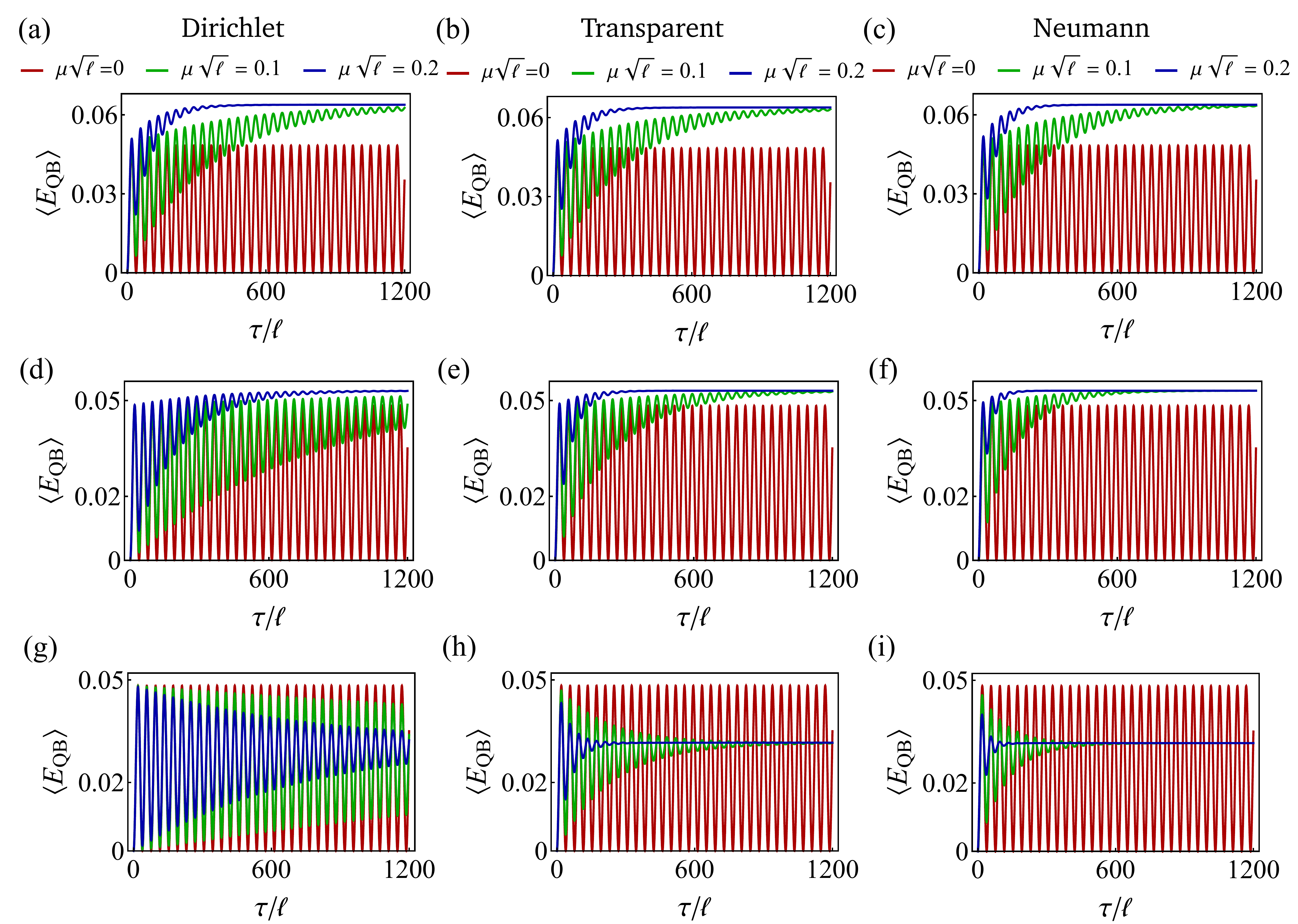}
\caption{Time evolution of the average energy stored in the QB $\langle\,E_\text{QB}(\tau)\rangle$ shown in Eq. \eqref{AEBD} for the decoherence coupling case ($\theta=0$). Three different boundary condition cases in BTZ spacetime are shown for various coupling strength $\mu\sqrt{\ell}$. Here we have taken
the energy level spacing $\Delta\ell=0.13$, the charging amplitude $A\ell=0.1$, and the mass $M=1$. The first, second, and third row panels have taken 
the relative distance of the QB in BTZ spacetime $r/r_h=1.001$, $r/r_h=1.1$ and $r/r_h=1.8$, respectively.}\label{fig9}
\end{figure*}

\begin{figure*}[t]
\centering
\includegraphics[width=0.9  \textwidth]{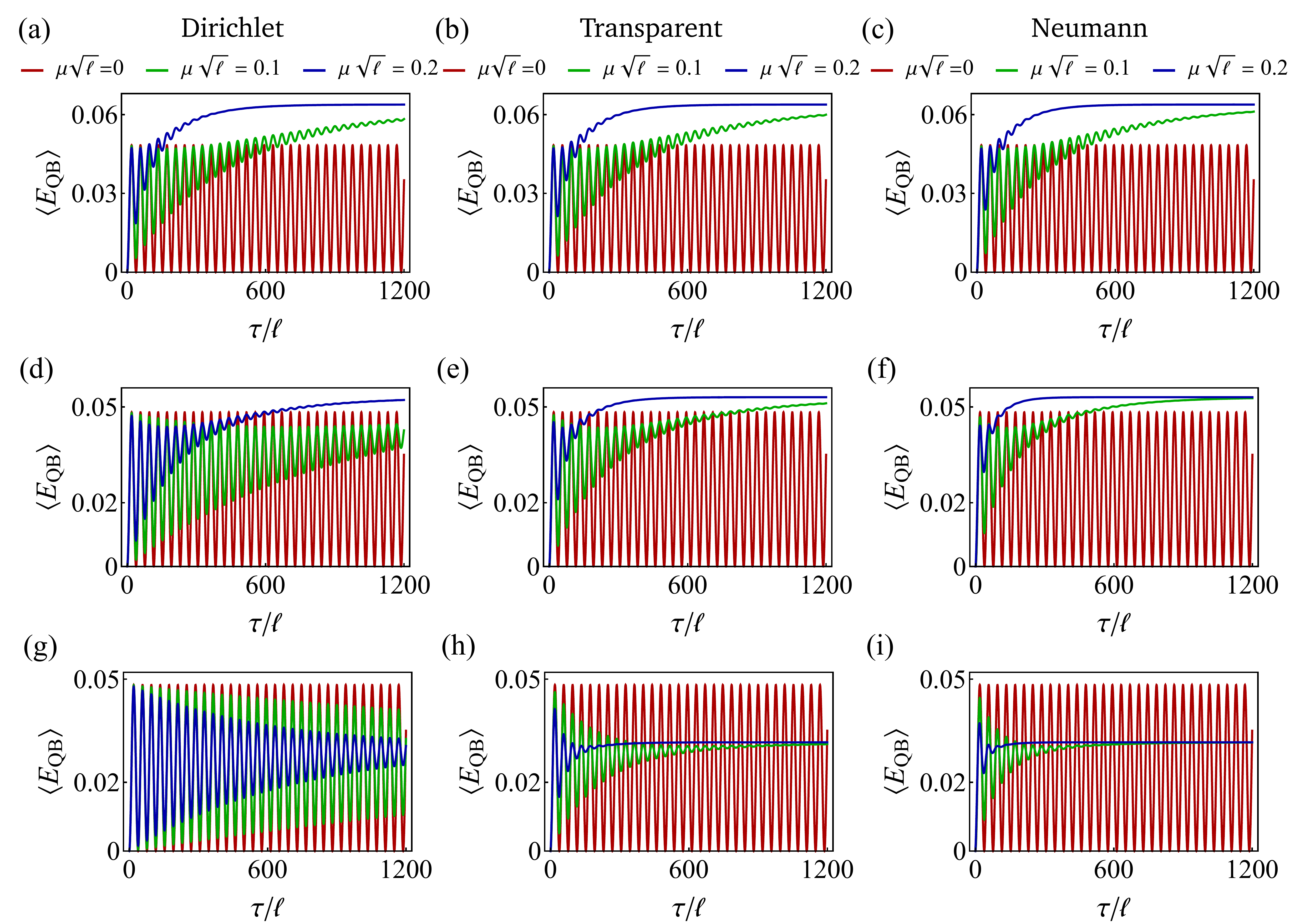}
\caption{Time evolution of the average energy stored in the QB $\langle\,E_\text{QB}(\tau)\rangle$ shown in Eq. \eqref{AEBD} for the pure dephasing coupling case ($\theta=\pi/2$). Three different boundary condition cases in BTZ spacetime are shown for various coupling strength $\mu\sqrt{\ell}$. Here we have taken
the energy level spacing $\Delta\ell=0.13$, the charging amplitude $A\ell=0.1$, and the mass $M=1$. The first, second, and third row panels have taken 
the relative distance of the QB in BTZ spacetime $r/r_h=1.001$, $r/r_h=1.1$ and $r/r_h=1.8$, respectively.}\label{fig10}
\end{figure*}

\subsubsection{$\Delta\ge\,A$ case}

To consider the charging dynamics for $\Delta\ge\,A$ case, we in Fig. \ref{fig9} plot the time evolution of the average energy stored in the QB $\langle\,E_\text{QB}(\tau)\rangle$ shown in Eq. \eqref{AEBD} for the decoherence coupling case ($\theta=0$). 
We fix the energy-level spacing of the QB $\Delta\ell=0.13$, the charging amplitude $A\ell=0.1$, and the mass $M=1$. Representative examples
of different distance of the QB (or local Hawking temperature) away from the horizon $r/r_h=1.001$ (Fig. \ref{fig9} (a), (b) and (c)), $r/r_h=1.1$ (Fig. \ref{fig9} (d), (e) and (f)), and $r/r_h=1.8$ (Fig. \ref{fig9} (g), (h) and (i)) have been chosen. The closed QB system driven by a static bias (see Eq. \eqref{EQB0}) is also reported as a reference limit (see red curves). It is found that, in the presence of dissipation, the charged average energy presents an oscillatory behavior in the early time, while it will finally approach a maximum for the long enough charging time 
if the local Hawking temperature is high. This maximum average energy depends on the local Hawking temperature, the charging amplitude $A\ell$, and 
the energy-level spacing of the QB $\Delta\ell$. Interestingly, unlike the $\Delta<A$ case, when $\Delta>A$ the maximum average energy might be more than
that for the closed QB case if the local Hawking temperature is high enough. Note that this advantage could occur for all the three different boundary condition cases, including the Dirichlet, the transparent, and the Neumann one. This means that the dissipation from vacuum field fluctuations in the BTZ spacetime might enhance the charging performance. In other words, through the interaction between the QB and the external quantum field one can in principle 
extract the energy from the vacuum fluctuations in curved space through dissipation in the charging process.
Furthermore, we can also find that for the Dirichlet boundary case the average energy reaches a 
maximum earlier with the increase of the local Hawking temperature. However, for the transparent boundary and the Neumann boundary cases when 
the local Hawking temperature becomes higher, the average energy reaches the corresponding maximum later. Further comparing the average-energy dynamics for these three boundary condition cases, we can find that under the same conditions the average energy for the Neumann boundary case would be the first to reach a maximum value,
then it is the transparent boundary case, and finally it is the Dirichlet one. We can also find that the stronger the coupling between the QB and the quantum field is,
the earlier the QB reaches the maximum average energy during the charging process. However, the maximum value of the average energy does not depend on
the coupling strength. This conclusion is valid for all the three boundary condition cases.

In Fig. \ref{fig10} we plot the time evolution of the average energy stored in the QB $\langle\,E_\text{QB}(\tau)\rangle$ shown in Eq. \eqref{AEBP} for the pure dephasing coupling case ($\theta=\pi/2$). We also fixed the energy-level spacing of the QB $\Delta\ell=0.13$, the charging amplitude $A\ell=0.1$, and the mass $M=1$. Representative examples of different distance of the QB (or local Hawking temperature) away from the horizon $r/r_h=1.001$ (Fig. \ref{fig10} (a), (b) and (c)), $r/r_h=1.1$ (Fig. \ref{fig10} (d), (e) and (f)), and $r/r_h=1.8$ (Fig. \ref{fig10} (g), (h) and (i)) have been chosen. We also choose the closed QB system driven by a static bias (see Eq. \eqref{EQB0}) as a reference limit (see red curves). Since the average energy for the dephasing coupling case has the similar formula to that for the decoherence coupling case, seen from 
Eqs. \eqref{AEBD} and \eqref{AEBP}, we can find that the average energy for the dephasing coupling case shares the similar dynamics to that for the decoherence coupling case 
discussed above. Specifically, although in the presence of dissipation the charged average energy presents an oscillatory behavior in the early time,
finally it will reach a constant if the charging time is long enough. This constant average energy denotes the maximum average energy of the QB during the charging process if the Hawking radiation is high enough, satisfying $T\ge\Omega/\ln\big[\frac{\Delta\Omega+\Delta^2-A^2}{\Delta\Omega+A^2-\Delta^2}\big]$ as discussed above. It is interesting that the maximum average energy for this case is stored more in the QB than that for the closed QB case. This result is quite different 
compared with  $\Delta<A$ scenario where the maximum average energy in the presence of dissipation would never be more than that for the closed QB case. For all the three boundary condition cases, this maximum average energy increases with the increase of the local Hawking temperature, while it does not depend on the coupling strength between the QB and the external quantum field. However, the coupling strength would affect the time at which the average energy reaches the maximum.
It is found that no matter for which boundary cases, the stronger the coupling is, the earlier the average energy reaches the maximum. Comparing the three different 
boundary cases, we can find for the Dirichlet boundary case the higher local Hawking temperature may lead to the earlier maximum average energy that is reached by the QB, while this is quite the opposite for the other two boundary cases.

We give a brief summary here: we can find that the dynamics of the average energy stored in the QB during the charging process depends on the quantum field boundary condition, the coupling form between the QB and external quantum field, the local Hawking temperature, the QB energy-level spacing, and the charging amplitude.
It is shown that under certain conditions the maximum average energy stored in an open QB might be more than that for the closed QB case. This advantage could occur for all the three different boundary condition cases, including the Dirichlet, the transparent, and the Neumann one. 
It means that the dissipation from vacuum field fluctuations in the BTZ spacetime might enhance the charging performance. 

\subsection{Direct comparison of $\langle\,E\rangle$ for different coupling form}
\begin{figure*}[t]
\centering
\includegraphics[width=0.9  \textwidth]{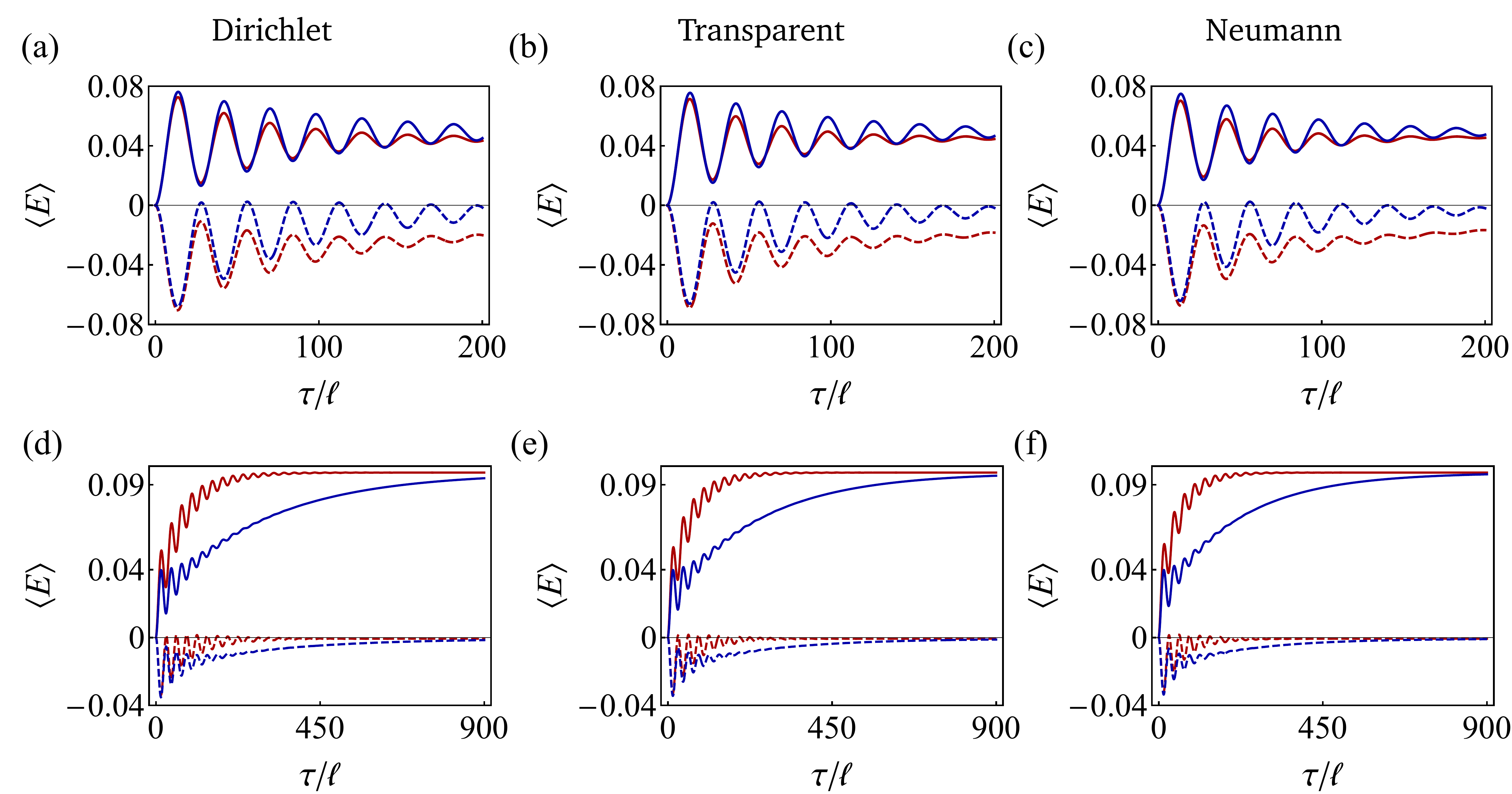}
\caption{Time evolution of the average energy variation $\langle\,E\rangle$ for $A\ell=0.2>\Delta\ell=0.1$ cases (a), (b), (c) and for $A\ell=0.1<\Delta\ell=0.2$ cases (d), (e), (f). Solid red and blue lines associated to the QB correspond to the case of decoherence ($\theta=0$) and pure dephasing ($\theta=\pi/2$) interaction with the quantum field, respectively; and dashed red and blue lines are associated to the charger. Here we have taken the mass $M=1$, the relative distance of the QB in BTZ spacetime $r/r_h=1.001$, the coupling strength $\mu\sqrt{\ell}=0.2$, respectively.}\label{fig11}
\end{figure*}
To better understand the charging performance in the presence of different dissipation, we directly compare two dissipative coupling schemes in Fig. \ref{fig11}. Two different driving amplitudes are considered, i.e., $A\ell=0.2>\Delta\ell=0.1$ and $A\ell=0.1<\Delta\ell=0.2$ with coupling strength $\mu\sqrt{\ell}=0.2$. For $A\ell>\Delta\ell$ case the plots show that in the case $\theta=\pi/2$ (the pure dephasing coupling case) it is possible to achieve a better charging of QB (higher values of the maxima) compared with the decoherence coupling case ($\theta=0$) in the finite time regime. This difference results from the different form of the dissipation rates in Eqs. \eqref{rates1}, \eqref{DR}, where we can find the dephasing coupling case 
is less affected by dissipation. Indeed, since we consider $r/r_h\rightarrow1$ (i.e., $T\rightarrow\infty$), Eqs. \eqref{AEBD} and \eqref{AEBP} reduce to
\begin{eqnarray}\label{AEBD1}
\langle\,E_\text{QB}(\tau)\rangle\approx\frac{\Delta}{2}\bigg\{1-\frac{\Delta^2}{\Omega^2}e^{-\Gamma^{\text{r}}_0\tau}-\frac{A^2}{\Omega^2}\cos(\Omega\tau)
e^{-\Gamma_0\tau}\bigg\},
\end{eqnarray}
and 
\begin{eqnarray}\label{AEBP1}
\langle\,E_\text{QB}(\tau)\rangle\approx\frac{\Delta}{2}\bigg\{1-\frac{\Delta^2}{\Omega^2}e^{-\Gamma^{\text{r}}_{\pi/2}\tau}-\frac{A^2}{\Omega^2}\cos(\Omega\tau)
e^{-\Gamma_{\pi/2}\tau}\bigg\},
\end{eqnarray}
respectively. We can numerically analyze the above conclusion from their corresponding rates,
e.g, for chosen parameters of Fig. \ref{fig11} (a), the corresponding
rates are $\Gamma^\text{r}_0\simeq0.00349\ell$, $\Gamma_0=0.0157\ell$ and $\Gamma^{\text{r}}_{\pi/2}\simeq0.01396\ell$, $\Gamma_{\pi/2}=0.01047\ell$, respectively. Since $\Gamma^\text{r}_0<\Gamma^{\text{r}}_{\pi/2}$, the smaller relaxation rate (characterized by $\Gamma^{\text{r}}_0$ and $\Gamma^{\text{r}}_{\pi/2}$) would lead to the smaller average energy with respect to the evolution time. Furthermore, the dephasing rate (characterized by $\Gamma_0$ and $\Gamma_{\pi/2}$) leads to the damping of oscillations (see in Figs. \ref{fig11} (a), (b) and (c)). Because $\Gamma_0>\Gamma_{\pi/2}$, the oscillatory amplitude of the average energy for the decoherence coupling case ($\theta=0$) might decay faster than the pure dephasing coupling case ($\theta=\pi/2$) due to the damping effect. 
As a result of that, in the finite time regime, it is possible to achieve a better charging of QB for the pure dephasing coupling case.
However, in the infinite time limit, both cases may achieve the same maximum average energy, given by $\langle E(\infty)\rangle=\Delta/2$.
Besides, notice that curves associated to the QB and to the charger are not specular (with respect to the $x$ axis), reflecting the fact that a given amount of energy is also dissipated into the reservoir. When $A\ell<\Delta\ell$, we find that the charging of the QB performances quite differently compared with that 
for the $A\ell>\Delta\ell$ case. Here we can find that although in the infinite charging time limit, both the decoherence coupling ($\theta=0$)  case and the pure dephasing coupling ($\theta=\pi/2$) case may achieve the same maximum average energy, while the decoherence coupling case may reach the maximum earlier than the pure dephasing coupling case. In this regard, the decoherence coupling case seems to be more efficient than the pure dephasing coupling case for charging protocol. 
Furthermore, we can also find that curves associated to the QB and to the charger are not specular (with respect to the $x$ axis) and the average-energy variation associated to the QB is more than that associated to the charging, reflecting the fact that the QB not only get the energy from the charge, but also can extract the energy from the reservoir. This means that one can extract the energy from the vacuum fluctuations of quantum field in curved spacetime through charging protocol.

\begin{figure*}[t]
\centering
\includegraphics[width=0.9 \textwidth]{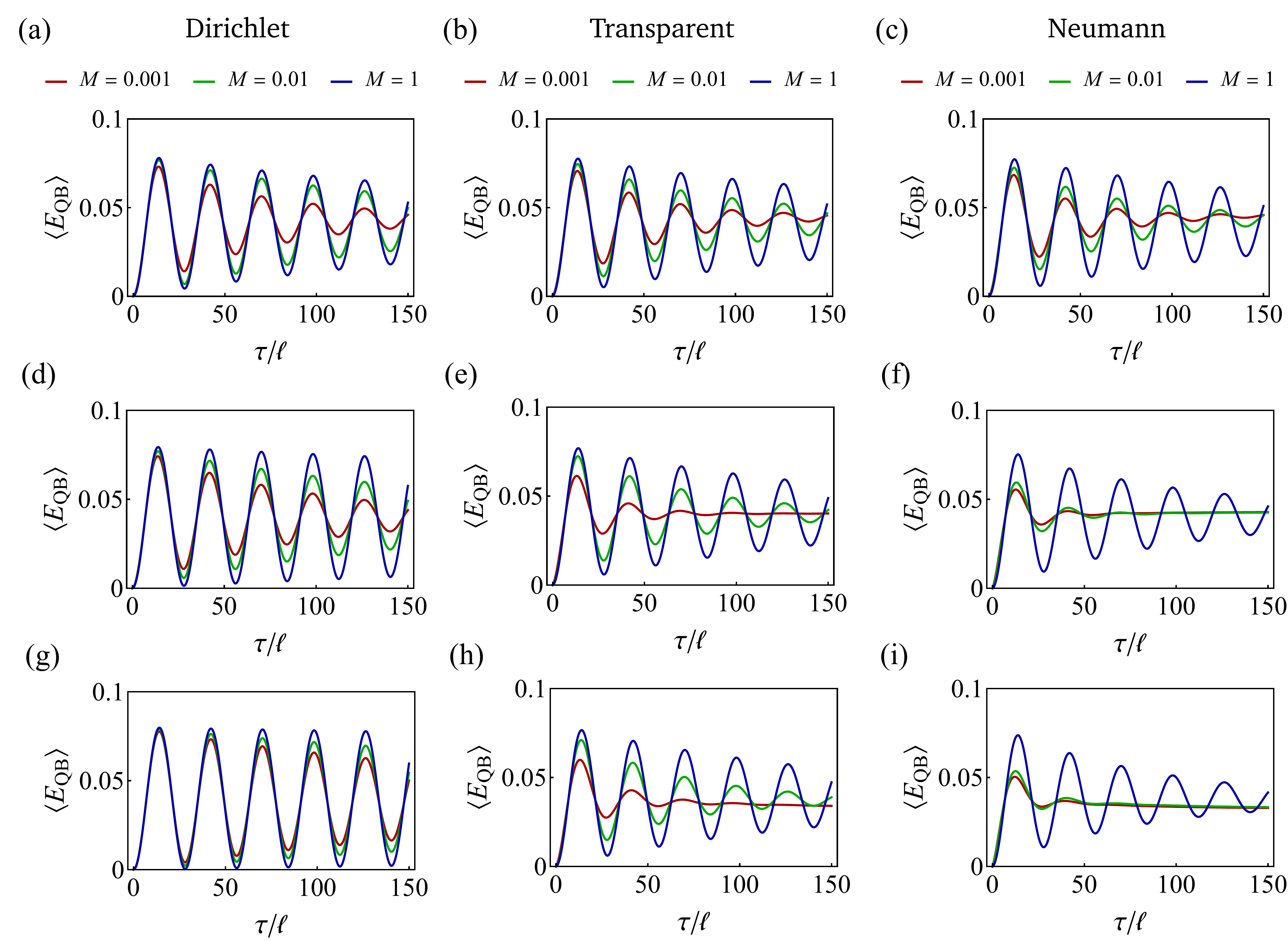}
\caption{Time evolution of the average energy stored in the QB $\langle\,E_\text{QB}(\tau)\rangle$ shown in Eq. \eqref{AEBD} for the decoherence coupling case with $\theta=0$. Three different boundary conditions in BTZ spacetime are shown for various mass of black hole  $M$. Here we have taken
the energy-level spacing $\Delta\ell=0.1$, the charging amplitude $A\ell=0.2$, and the coupling strength $\mu\sqrt{\ell}=0.1$. The first, second, and third row panels have taken 
the relative distance of the QB in BTZ spacetime $r/r_h=1.001$, $r/r_h=1.2$ and $r/r_h=1.8$, respectively.}\label{fig3M}
\end{figure*}

\subsection{Effect of mass of BTZ black hole on average energy}

We have studied the dynamics of the average energy stored in an open QB that is located in the BTZ spacetime in the above sections. 
Where we fixed the mass of the black hole and mainly considered how the effective local black hole temperature,  the driving amplitude, the boundary conditions on quantum field, and different dissipative couplings affect the relevant dynamics. Since small-mass black holes can produce interesting effects in relativistic quantum information, such as Fisher information \cite{FIBH}, it is interesting to see what happens for the charging performances of a QB when the mass of black hole is small.

\subsubsection{$\Delta<A$ case}

\begin{figure*}[t]
\centering
\includegraphics[width=0.9 \textwidth]{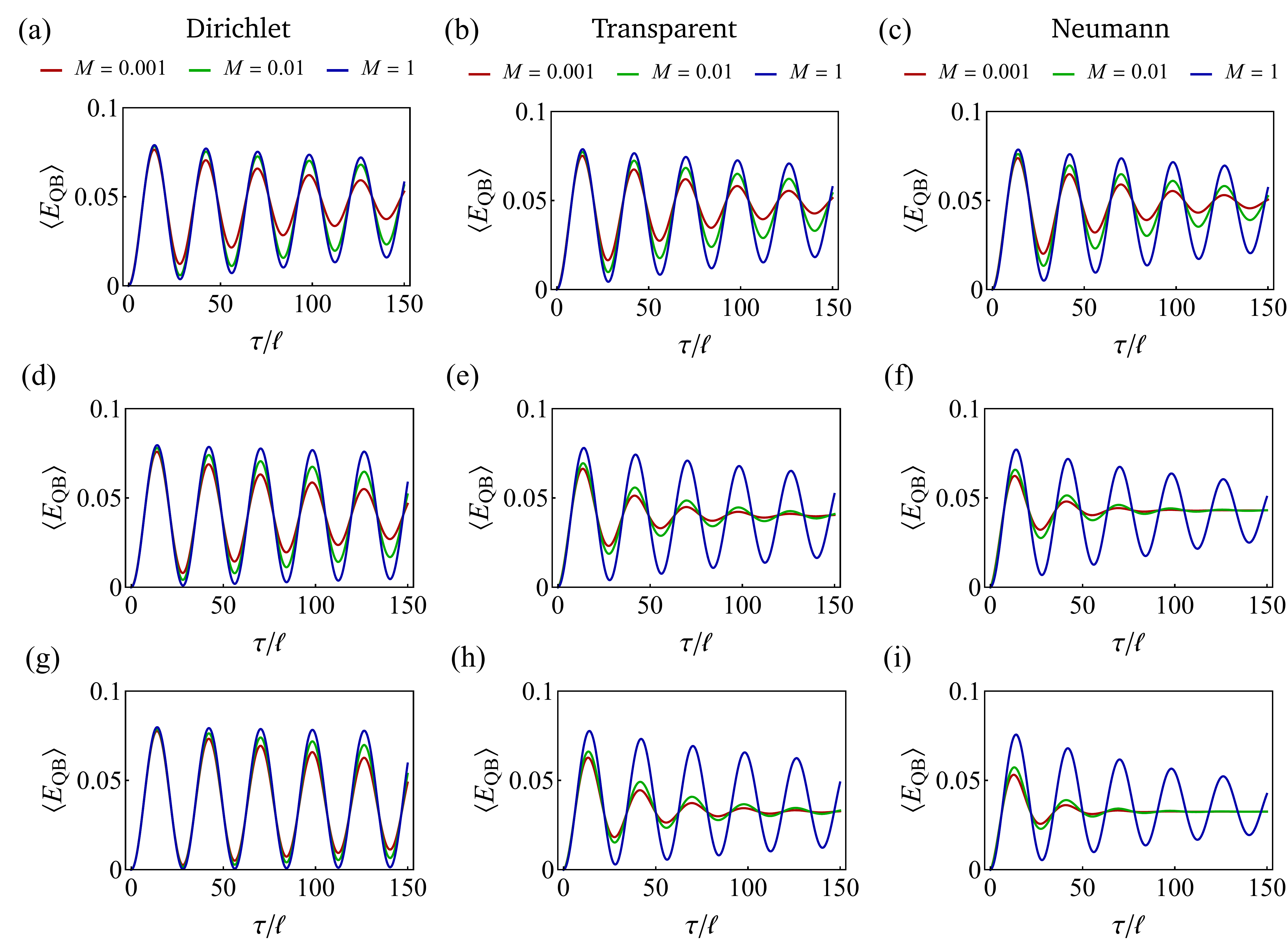}
\caption{Time evolution of the average energy stored in the QB $\langle\,E_\text{QB}(\tau)\rangle$ shown in Eq. \eqref{AEBP} for the pure dephasing coupling case with $\theta=\pi/2$. Three different boundary conditions in BTZ spacetime are shown for various mass of black hole $M$. Here we have taken
the energy level spacing $\Delta\ell=0.1$, the charging amplitude $A\ell=0.2$, and the coupling strength $\mu\sqrt{\ell}=0.1$. The first, second, and third row panels have taken the relative distance of the QB in BTZ spacetime $r/r_h=1.001$, $r/r_h=1.2$ and $r/r_h=1.8$, respectively.}\label{fig4M}
\end{figure*}

In Fig. \ref{fig3M} we show the time evolution of the average energy stored in the QB $\langle\,E_\text{QB}(\tau)\rangle$ shown in Eq. \eqref{AEBD} for the decoherence coupling case 
($\theta=0$ case) with different black hole masses $M=0.001$, $M=0.01$ and $M=1$. We fixed the energy-level spacing of the QB $\Delta\ell=0.1$, the charging amplitude $A\ell=0.2$, and the coupling strength $\mu\sqrt{\ell}=0.1$. Representative examples
of different distance of the QB (or local Hawking temperature) away from the horizon $r/r_h=1.001$ (Fig. \ref{fig3M} (a), (b) and (c)), $r/r_h=1.2$ (Fig. \ref{fig3M} (d), (e) and (f)), and $r/r_h=1.8$ (Fig. \ref{fig3M} (g), (h) and (i)) have been chosen. It is found that, in the presence of dissipation, the charged average energy presents a damped oscillatory behavior, whose amplitude is modulated by the exponential decay dictated by the incoherent relaxation and dephasing rates given in Eqs. \eqref{rates1}. When fixing the effective local Hawking temperature, we can find that with the increase of the black hole mass $M$, the damping strength 
is degraded. It means that the average energy stored in the QB degrades faster to its asymptotic limit \eqref{QBE} for smaller black hole mass. However, 
the smaller the black hole mass is, the less the maximum average energy stored in the QB during the charging process is. Therefore, if the local Hawking temperature 
is fixed, the mass of black hole can not lead to qualitatively different dynamical behavior of the average energy stored in the QB, this is different from
the Fisher information case \cite{FIBH} where small black hole mass can induce qualitative difference of the Fisher information dynamics. We can understand this 
from the expression of the average energy shown in Eq. \eqref{AEBD}. If the local Hawking temperature is fixed, the mass of black hole can only change 
the incoherent relaxation and dephasing rates given in Eqs. \eqref{rates1}, thus can only change the damping strength. Besides, no matter for which 
black hole mass, the local Hawking temperature can enhance the damping of the dynamics of the average energy that is stored in the QB for the  Dirichlet boundary case, while for the transparent and the Neumann boundary cases, the local Hawking temperature would degrade the damping of the dynamics of the average energy.
Under the same condition, we can see that the Neumann boundary condition causes stronger damping than the transparent one, while the transparent one induces stronger damping than the Dirichlet one.

In Fig. \ref{fig4M} we show the time evolution of the average energy stored in the QB $\langle\,E_\text{QB}(\tau)\rangle$ shown in Eq. \eqref{AEBP} for the pure dephasing coupling case ($\theta=\pi/2$) with different black hole masses $M=0.001$, $M=0.01$ and $M=1$. 
As comparison, we fixed the same other parameters as that in Fig. \ref{fig3M}. We find that in the presence of dissipation all curves present a damped oscillatory behaviour, whose amplitude is modulated by the exponential decay dictated by the incoherent relaxation and dephasing rates given in Eqs. \eqref{DR}. With the decrease of the black hole mass, the dynamics of the average energy presents a stronger damping. Which means that the smaller the black hole mass is,
the faster the average energy approaches its asymptotical limit shown in \eqref{QBE}. However, the smaller the black hole mass is, the smaller of the maximum 
average energy during the charging process is.

From the above analysis, we can find that the mass of black hole just changes the incoherent relaxation and dephasing rates shown in the Eqs. \eqref{rates1} and  \eqref{DR}, thus only modifies damping strength of the dynamics of the average energy stored in the QB. Actually, one can arrive at this conclusion directly 
from the expressions of the average energy shown in \eqref{AEBD} and \eqref{AEBP}.

 \begin{figure*}[t]
\centering
\includegraphics[width=0.9 \textwidth]{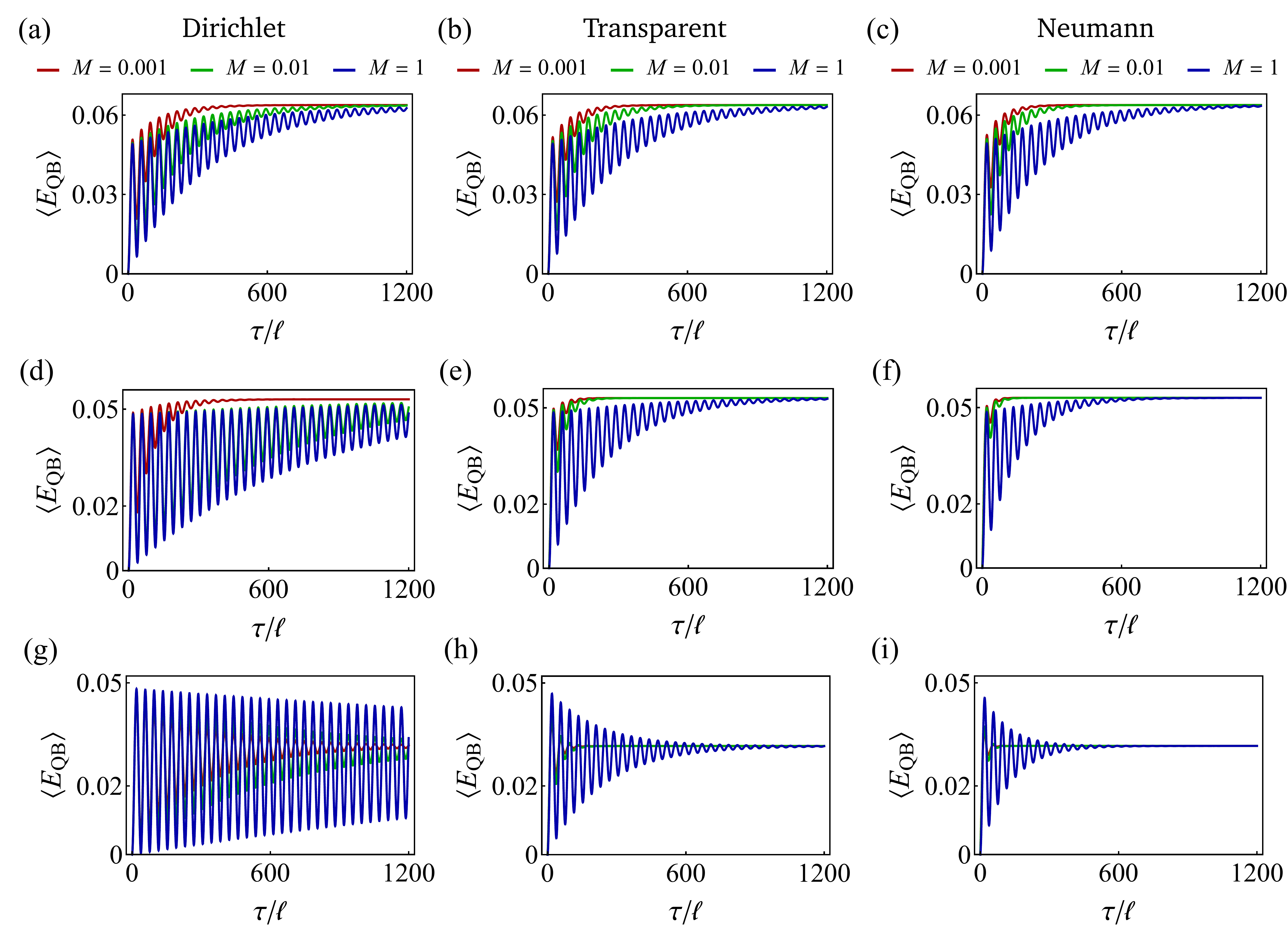}
\caption{Time evolution of the average energy stored in the QB $\langle\,E_\text{QB}(\tau)\rangle$ shown in Eq. \eqref{AEBD} for the decoherence coupling case ($\theta=0$). Three different boundary condition cases in BTZ spacetime are shown for various black hole mass $M=0.001$, $M=0.01$, and $M=1$. 
Here we have taken the energy level spacing $\Delta\ell=0.13$, the charging amplitude $A\ell=0.1$, and the coupling strength $\mu\sqrt{\ell}=0.1$. 
The first, second, and third row panels have taken 
the relative distance of the QB in BTZ spacetime $r/r_h=1.001$, $r/r_h=1.1$ and $r/r_h=1.8$, respectively.}\label{fig9M}
\end{figure*}

\begin{figure*}[t]
\centering
\includegraphics[width=0.9  \textwidth]{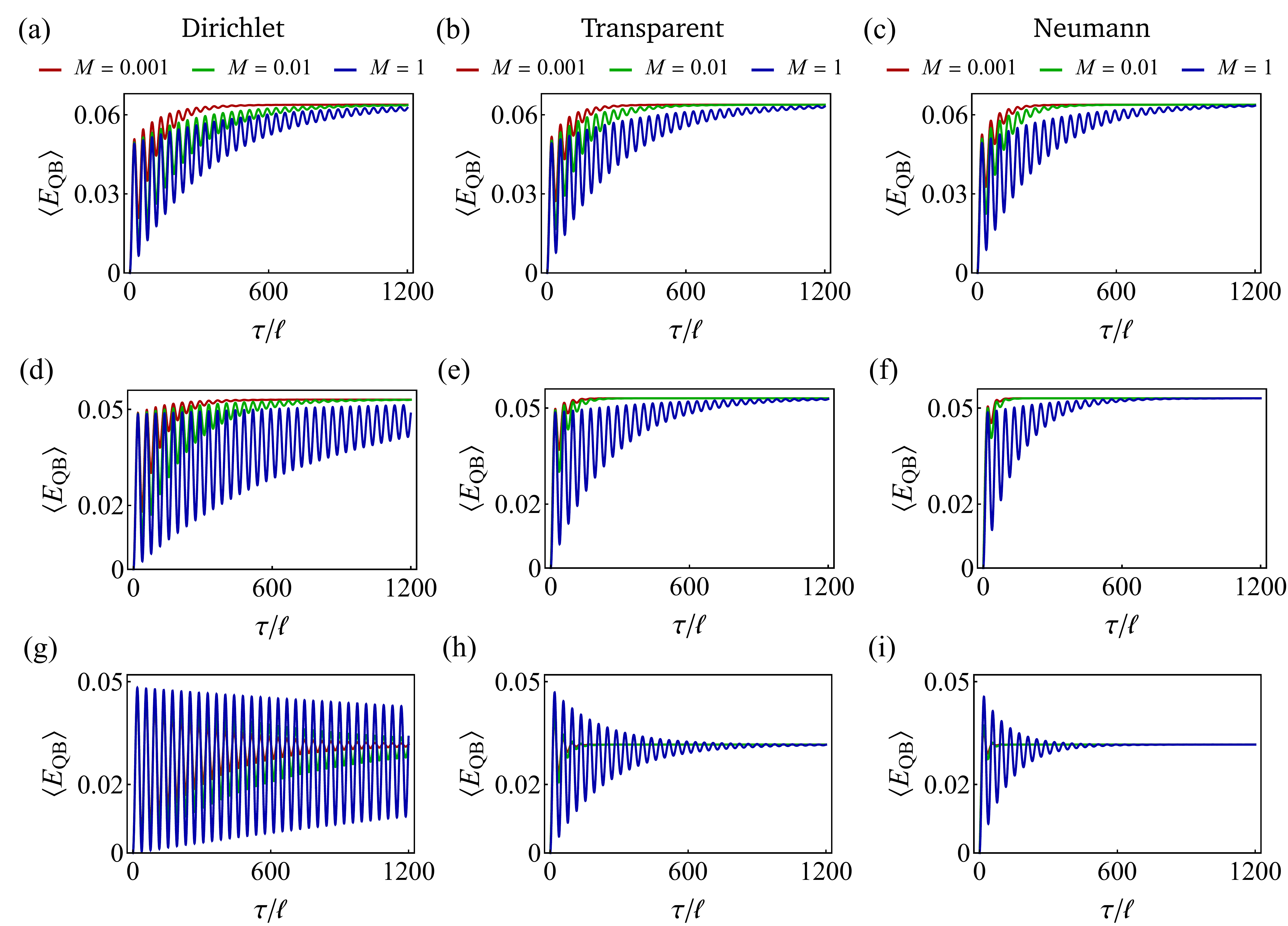}
\caption{Time evolution of the average energy stored in the QB $\langle\,E_\text{QB}(\tau)\rangle$ shown in Eq. \eqref{AEBD} for the pure dephasing coupling case ($\theta=\pi/2$). Three different boundary condition cases in BTZ spacetime are shown for various black hole mass $M=0.001$, $M=0.01$, and $M=1$. 
Here we have taken
the energy level spacing $\Delta\ell=0.13$, the charging amplitude $A\ell=0.1$, and the coupling strength $\mu\sqrt{\ell}=0.1$. The first, second, and third row panels have taken 
the relative distance of the QB in BTZ spacetime $r/r_h=1.001$, $r/r_h=1.1$ and $r/r_h=1.8$, respectively.}\label{fig10M}
\end{figure*}

\subsubsection{$\Delta\ge\,A$ case}
In the above section we have analyzed how the black hole mass $M$ affects the average energy stored in the QB for the $\Delta<A$ case. Here we will 
discuss what happens for the average energy of QB when $\Delta\ge\,A$ in the following. 

We fixed the coupled strength $\mu\sqrt{\ell}=0.1$, the energy level spacing $\Delta\ell=0.13$, effective local Hawking temperature, and the charging amplitude $A\ell=0.1$, and plotted the dynamics of the average energy of QB with different black hole mass $M=0.001$, $M=0.01$, and $M=1$ 
in Figs. \ref{fig9M} and \ref{fig10M}, which denote the decoherence coupling case ($\theta=0$) and the pure dephasing coupling case ($\theta=\pi/2$), respectively.
We can find that when we fix the other parameters, the mass of the black hole can not lead to qualitatively different dynamical behavior of the average energy.
All the average energy presents oscillatory behavior with respect to the evolution time, and finally approaches to an asymptotical value at the infinite 
time limit. We can find that when the local Hawking temperature is fixed, this asymptotical value of the average energy is the same for all the black hole mass cases.
However, we can find that the mass of the black hole could affect the speed at which the average energy evolves to its asymptotical value shown in \eqref{QBE}.
The smaller the black hole mass is, the faster the average energy evolves to its asymptotical value.

As discussed for the $\Delta<A$ case above, we can arrive at the relevant conclusion directly from the expression of the average energy of the QB shown in  
the Eqs. \eqref{AEBD} and  \eqref{AEBP}. When the local Hawking temperature is fixed, the black hole mass can only affect the incoherent relaxation and dephasing rates shown in the Eqs. \eqref{rates1} and  \eqref{DR}, thus can not lead to qualitatively behavior. Note that this result of QB here is quite different from 
that of the Fisher information when the mass is changed shown in Ref. \cite{FIBH}, where the behavior of the Fisher information can present qualitatively
different behavior when the black hole mass is changed.

\subsection{Charging stability}
Another important question regards how long a QB can retain a given amount of energy that has been stored during a charging process, in the presence of dissipation.
To answer this question, we will inspect the following protocol. At $\tau\ge0$ we assume the QB is charged by the external charger which has the amplitude $A$, and this charging process lasts to $\tau_c$. Then we let the charger be switched off, i.e., $A=0$, for $\tau>\tau_c$. For this kind of protocol, there is in general an energy cost 
$\delta\,E_\text{SW}=-\langle\,H_\text{C}\rangle$ (evaluated at the time $\tau_c$) associated to the switching operation. Therefore, in a fully quantum case, as noted in Refs. \cite{PhysRevB.98.205423, Chetcuti_2020}, one has to determine whether there is enough energy in the system (the charger) to compensate this additional cost. However, this is not problem in our model, since our QB is charged by a classical source which can be assumed to be big enough to provide the energy required to implement the switch without affecting the performance of the QB. For a closed system, the amount of energy stored in the QB until the time $\tau_c$ will remain constant at later times. 
However, unlike the closed one, a more realistic model---open QB as a result of the interaction with environment will inevitably dissipate part of its energy into the thermal bath. Therefore, it is quite important to quantify the amount of energy retained in the QB, after the charging field has been switched off. 
To model this protocol, we consider that at $\tau=0$ the QB is initially prepared in a given ground state. The system then evolves under the action of the charger with
amplitude $A$ until the time $\tau_c$. At $\tau=\tau_c$ the reduced density matrix of the QB is $\tilde{\rho}_\text{QB}(\tau_c)=\frac{1}{2}(\mathrm{I}+\vec{\mathbf{r}}(\tau_c)\cdot\vec{\sigma})$ with the state parameters given in \eqref{Rotation-S}. Correspondingly, the average energy stored in the QB is given by 
\begin{eqnarray}\label{AEBD1}
\nonumber
\langle\,E_\text{QB}(\tau_c)\rangle_\text{DC}&=&\frac{\Delta}{2}\bigg\{1-\frac{\Delta^2}{\Omega^2}e^{-\Gamma^{\text{r}}_0\tau_c}-
\frac{\Delta}{\Omega}\bigg(\frac{e^{\Omega/T}-1}{e^{\Omega/T}+1}\bigg)\big(1-e^{-\Gamma^{\text{r}}_0\tau_c}\big)-\frac{A^2}{\Omega^2}\cos(\Omega\tau_c)
e^{-\Gamma_0\tau_c}\bigg\},
\\
\end{eqnarray}
for the decoherence coupling case, and
\begin{eqnarray}\label{AEBP1}
\nonumber
\langle\,E_\text{QB}(\tau_c)\rangle_\text{DP}&=&\frac{\Delta}{2}\bigg\{1-\frac{\Delta^2}{\Omega^2}e^{-\Gamma^{\text{r}}_{\pi/2}\tau_c}-
\frac{\Delta}{\Omega}\bigg(\frac{e^{\Omega/T}-1}{e^{\Omega/T}+1}\bigg)\big(1-e^{-\Gamma^{\text{r}}_{\pi/2}\tau_c}\big)-\frac{A^2}{\Omega^2}\cos(\Omega\tau_c)
e^{-\Gamma_{\pi/2}\tau_c}\bigg\},
\\
\end{eqnarray}
for the pure dephasing coupling case, respectively. When $\tau\ge\tau_c$ the charger is switched off (with the amplitude $A=0$) and therefore 
this operation may lead to a different dynamics for the QB. In this scenario, the evolution of  the QB can also be described with Eq. \eqref{TES}
while the state parameters in Eq. \eqref{Rotation-S} are different from the $A\neq0$ case discussed above. Besides, in this case, the effective initial state 
of the QB can be described in terms of the reduced density matrix $\tilde{\rho}_\text{QB}(\tau_c)$. Specifically, 
as shown in \eqref{T-D} the associated parameters for this case can be written as 
\begin{eqnarray}\label{T-D1}
\nonumber
\gamma_+&=&\frac{\mu^2}{2}\bigg(\frac{1}{e^{\Omega/T}+1}\bigg)\sum^{n=\infty}_{n=-\infty}
\bigg[P_{-\frac{1}{2}+i\frac{\Omega}{2\pi\,T}}(\cosh\alpha_n)-\zeta\,P_{-\frac{1}{2}+i\frac{\Omega}{2\pi\,T}}(\cosh\beta_n)\bigg],
\\          \nonumber
\gamma_-&=&\frac{\mu^2}{2}\bigg(\frac{1}{e^{-\Omega/T}+1}\bigg)\sum^{n=\infty}_{n=-\infty}
\bigg[P_{-\frac{1}{2}-i\frac{\Omega}{2\pi\,T}}(\cosh\alpha_n)-\zeta\,P_{-\frac{1}{2}-i\frac{\Omega}{2\pi\,T}}(\cosh\beta_n)\bigg],
\\
\gamma_z&=&0,
\end{eqnarray}
for the decoherence coupling case, and
\begin{eqnarray}\label{T-D2}
\nonumber
\gamma_+&=&0,~~~~~~\gamma_-=0,
\\
\gamma_z&=&\frac{\mu^2}{4}\sum^{n=\infty}_{n=-\infty}
\bigg[P_{-\frac{1}{2}}(\cosh\alpha_n)-\zeta\,P_{-\frac{1}{2}}(\cosh\beta_n)\bigg],
\end{eqnarray}
for the pure dephasing case, respectively. Besides, when $\tau>\tau_\text{c}$ the evolution of the average energy stored in the QB subjected to external quantum field vacuum fluctuations reads 
\begin{eqnarray}
\langle\,E(\tau)\rangle=\frac{\Delta}{2}\bigg\{1+\bigg[\frac{2\langle\,E_\text{QB}(\tau_c)\rangle_\text{DC}}{\Delta}-1\bigg]e^{-\Gamma^\text{r}_0\tau}-
\bigg(\frac{e^{\Omega/T}-1}{e^{\Omega/T}+1}\bigg)\big(1-e^{-\Gamma^{\text{r}}_0\tau}\big)\bigg\},
\end{eqnarray}
for the decoherence coupling case.
Here, $\Gamma^\text{r}_0=\gamma_++\gamma_-$ with $\gamma_\pm$ being in \eqref{T-D1}.
For the pure dephasing coupling case, we can find that the average energy stored in the QB after switching off the charging does not change, i.e., $\langle\,E(\tau)\rangle=\langle\,E_\text{QB}(\tau_c)\rangle_\text{DP}$.

\begin{figure*}[t]
\centering
\includegraphics[width=0.9  \textwidth]{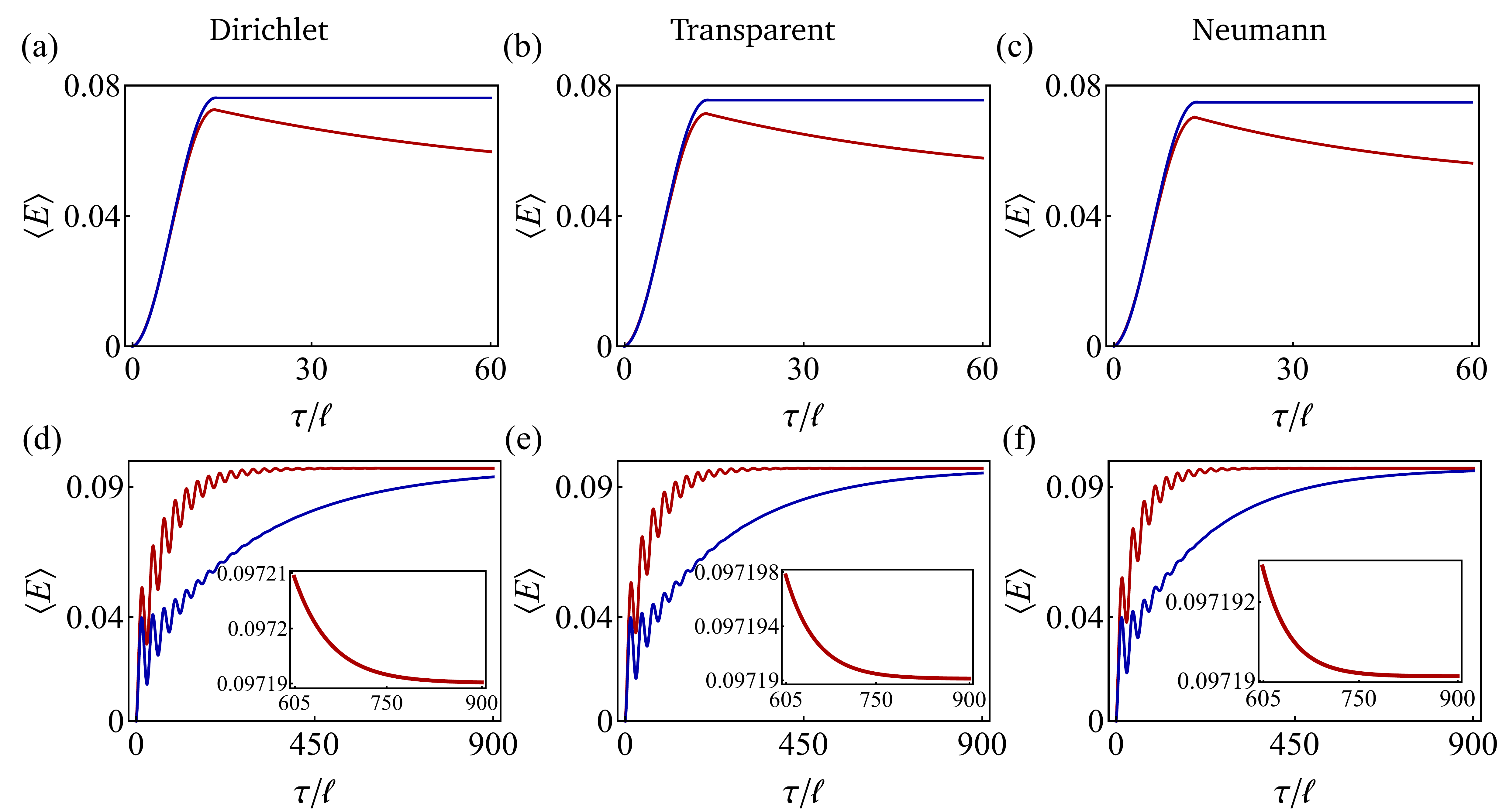}
\caption{Stability of average energy in the presence of dissipation: (a), (b) and (c) for the $A\ell=0.2>\Delta\ell=0.1$ case, while (d), (e), and (f) for the 
$A\ell=0.1<\Delta\ell=0.2$ case. The red curves denote the decoherence coupling case ($\theta=0$), while the blue curves  denote the pure dephasing coupling case ($\theta=\pi/2$). Note that the insets denote the detailed decay for the decoherence coupling case, while the pure dephasing coupling corresponds to the QB effectively decoupled from the external quantum field when the driving field is switched off. Here we have taken the coupling strength $\mu\sqrt{\ell}=0.2$, the mass $M=1$, and the relative distance of the QB in BTZ spacetime $r/r_h=1.001$, respectively.}\label{fig12}
\end{figure*}

In Fig. \ref{fig12} we show the time evolution of the average energy stored in the QB under this protocol (switching-off the charging ($A=0$) at time $\tau_c$). We can find that when $A>\Delta$ ( Fig. \ref{fig12} (a), (b), and (c)) for the decoherence coupling case the stored average energy, after the first maximum (at time $\tau_c$) starts to decrease. However, for the pure dephasing coupling case the stored average energy, after the first maximum, presents a flat behaviour, as a result of the effective decoupling of the QB from the external quantum field (resembling the dynamics of a closed system). Besides, we can find the blue curves is above the red ones, which means that more energy can be stored for the pure dephasing coupling case, reflecting a better charging performance compared to the  decoherence coupling case. This result is valid for all the three different boundary condition cases. When $A<\Delta$ (Fig. \ref{fig12} (d), (e), and (f)), we can find that the maximum average energy for both the decoherence coupling and the pure dephasing coupling cases is reached only when the system approaches to its equilibrium state. However, compared with the pure dephasing coupling case
the average energy stored in the QB for the decoherence coupling case reaches its equilibrium station earlier. We can choose an appropriate time (also denoted as $\tau_c$) to switch off the driving field, at which the corresponding average energy stored in the QB for the pure decoherence coupling case is high enough, so is almost the same as the maximum one. After this value the stored average energy starts to decrease (seen from the insets), while this degradation is quite trivial. For the pure dephasing coupling case, at that moment (at the chosen $\tau_c$) the average energy stored in the QB would not reach the maximum and is always smaller than that for the decoherence coupling case. 
This result is valid for all the three different boundary condition cases. In this regard, when $A<\Delta$ more energy can be stored for the decoherence coupling case
in a finite time regime, reflecting a better charging performance compared to the pure dephasing coupling case.

\section{Quantum battery in AdS spacetime}\label{section4}
Since the BTZ black hole spacetime can be obtained from identifications of $(2+1)$-dimensional AdS spacetime, so does the relevant Wightman function of quantum field shown above, it is interesting to see whether there is difference of the charging performance for these two spacetimes. In our case, $(2+1)$-dimensional AdS with cosmological constant $\Lambda=-1/\ell^2$ can be expressed as the hyperboloid $X^2_1+X^2_2-T^2_1-T^2_2=-\ell^2$, embedded in the $(2+2)$-dimensional flat spacetime $dS^2=dX^2_1+dX^2_2-dT^2_1-dT^2_2$, with $\ell$ being the AdS-length. We will consider 
the AdS-Rinder metric 
\begin{eqnarray}
ds^2=-\bigg(\frac{r^2}{\ell^2}-1\bigg)dt^2+\bigg(\frac{r^2}{\ell^2}-1\bigg)^{-1}dr^2+r^2d\phi^2,
\end{eqnarray}
which is obtained via the transformations:
\begin{eqnarray}
\nonumber
T_1=\ell\sqrt{\frac{r^2}{\ell^2}}\cosh\phi,~~X_1=\ell\sqrt{\frac{r^2}{\ell^2}}\sinh\phi,~~
T_2=\ell\sqrt{\frac{r^2}{\ell^2}-1}\sinh\frac{t}{\ell},~~X_2=\ell\sqrt{\frac{r^2}{\ell^2}-1}\cosh\frac{t}{\ell}.
\\
\end{eqnarray}
Note that an accelerate horizon presents at $r_\text{h}=\ell$. Correspondingly, the vacuum Wightman function for a (massless) conformally coupled scalar field in $(2+1)$-dimensional AdS reads 
\begin{eqnarray}\label{AdS-W}
G^+_\text{AdS}(x, x^\prime)=\frac{1}{4\pi\sqrt{2}\ell}\bigg(\frac{1}{\sqrt{\sigma(x, x^\prime)}}-\frac{\zeta}{\sqrt{\sigma(x, x^\prime)+2}}\bigg),
\end{eqnarray}
where $\sigma(x, x^\prime)=\frac{1}{2\ell^2}\big[(X_1-X^\prime_1)^2-(T_1-T^\prime_1)^2+(X_2-X^\prime_2)^2-(T_2-T^\prime_2)^2\big]$ denotes 
the square distance between $x$ and $x^\prime$ in the embedding space. Similarly, $\zeta\in\{1, 0, -1\}$ respectively specifies the the Dirichlet $(\zeta=1)$,
transparent $(\zeta=0)$, and Neumann $(\zeta=-1)$ boundary conditions satisfied by the field at spatial infinity \cite{PhysRevD.18.3565}.

Similar to the BTZ black hole case, we also assume that the QB is spatially fixed at a constant $r$ in the AdS spacetime such that $\delta\phi=0$, 
$\Delta\tau=\tau-\tau^\prime=\sqrt{-g_{00}}\Delta\tau$. In this case, inserting the AdS Wightman function from  \eqref{AdS-W} into the QB
state parameter \eqref{SP}, we can find 
\begin{eqnarray}
\nonumber
\gamma_+&=&\frac{\mu^2\cos^2(\theta-\Theta)}{2}\bigg(\frac{1}{e^{\Omega/T}+1}\bigg)\bigg[1-\zeta P_{-\frac{1}{2}+i\frac{\Omega}{2\pi T}}
(1+8\pi^2\ell^2T^2)\bigg],
\\               \nonumber
\gamma_-&=&\frac{\mu^2\cos^2(\theta-\Theta)}{2}\bigg(\frac{1}{e^{-\Omega/T}+1}\bigg)\bigg[1-\zeta P_{-\frac{1}{2}-i\frac{\Omega}{2\pi T}}
(1+8\pi^2\ell^2T^2)\bigg],
\\
\gamma_z&=&\frac{\mu^2\sin^2(\theta-\Theta)}{4}\bigg[1-\zeta P_{-\frac{1}{2}}(1+8\pi^2\ell^2T^2)\bigg],
\end{eqnarray}
where $P_\nu(x)$ denotes the associated Legendre function of the first kind. Therefore, one can find the ratio between the excitation rate and the deexcitation rate
is $\gamma_+/\gamma_-=e^{-\Omega/T}$. It means the QB will feel as if it were immersed into a thermal bath with the temperature 
$T=r_\text{h}/(2\pi\ell\sqrt{r^2-r^2_\text{h}})$, that is defined as the local KMS temperature of the field in AdS spacetime.
Note that this temperature $T=r_\text{h}/(2\pi\ell\sqrt{r^2-r^2_\text{h}})$ actually is the temperature felt by an accelerated observer in $\mathrm{AdS}_3$ spacetime  and the magnitude of the acceleration is given by $|a|=\frac{r}{\ell\sqrt{r^2-\ell^2}}$  \cite{HENDERSON2020135732}, where $r$ denotes the location of the observer. 
The minimum acceleration is $|a|=1/\ell$ which happens near $r=\infty$, while $|a|\rightarrow\infty$ when $r\rightarrow\ell$.
It is demonstrated in Ref. \cite{Jennings_2010}
that there exists a critical acceleration, $a_c=1/\ell$, in $\mathrm{AdS}_3$ spacetime, and only the observer with a constant super-critical acceleration 
$a$ (i.e., $a>1/\ell$) can register the quasi-thermal response with a temperature equal to $T=\sqrt{a^2-\ell^{-2}}/(2\pi)$. We will consider this kind of acceleration 
hereafter.

Repeat the same calculations as that for the BTZ black hole case shown above, we can also find that for the case of decoherence coupling ($\theta=0$ 
in Eq. \eqref{Interaction-H}) the corresponding average energy associated to the QB in the AdS spacetime is 
\begin{eqnarray}\label{DEAEAdS}
\langle E_\text{QB}(\tau)\rangle=\frac{\Delta}{2}\bigg\{1-\frac{\Delta^2}{\Omega^2}e^{-\Gamma^\text{r}_0\tau}-\frac{\Delta}{\Omega}
\bigg(\frac{e^{\Omega/T}-1}{e^{\Omega/T}+1}\bigg)(1-e^{-\Gamma^\text{r}_0\tau})-\frac{A^2}{\Omega^2}\cos(\Omega\tau)e^{-\Gamma_0\tau}\bigg\},
\end{eqnarray}
where 
\begin{eqnarray}
\nonumber
\Gamma^\text{r}_0&=&\frac{\mu^2\Delta^2}{2\Omega^2}\big[1-\zeta P_{-\frac{1}{2}+i\frac{\Omega}{2\pi T}}(1+8\pi^2\ell^2T^2)\big],
\\
\Gamma_0&=&\frac{1}{2}\Gamma^\text{r}_0+\frac{\mu^2A^2}{2\Omega^2}\big[1-\zeta P_{-\frac{1}{2}}(1+8\pi^2\ell^2T^2)\big].
\end{eqnarray}
While for the pure dephasing coupling case ($\theta=\pi/2$ in Eq. \eqref{Interaction-H}), the corresponding average energy associated to the QB is found to be
\begin{eqnarray}\label{DAEAdS}
\langle E_\text{QB}(\tau)\rangle=\frac{\Delta}{2}\bigg\{1-\frac{\Delta^2}{\Omega^2}e^{-\Gamma^\text{r}_{\pi/2}\tau}-\frac{\Delta}{\Omega}
\bigg(\frac{e^{\Omega/T}-1}{e^{\Omega/T}+1}\bigg)(1-e^{-\Gamma^\text{r}_{\pi/2}\tau})-\frac{A^2}{\Omega^2}\cos(\Omega\tau)e^{-\Gamma_{\pi/2}\tau}\bigg\},
\end{eqnarray}
where 
\begin{eqnarray}
\nonumber
\Gamma^\text{r}_{\pi/2}&=&\frac{\mu^2A^2}{2\Omega^2}\big[1-\zeta P_{-\frac{1}{2}+i\frac{\Omega}{2\pi T}}(1+8\pi^2\ell^2T^2)\big],
\\
\Gamma_{\pi/2}&=&\frac{1}{2}\Gamma^\text{r}_{\pi/2}+\frac{\mu^2\Delta^2}{2\Omega^2}\big[1-\zeta P_{-\frac{1}{2}}(1+8\pi^2\ell^2T^2)\big].
\end{eqnarray}
Comparing the average energies in Eqs. \eqref{DEAEAdS} and \eqref{DAEAdS} (for AdS spacetime case) with that in Eqs. \eqref{AEBD} 
and \eqref{AEBP} (for the BTZ black hole case), we can find that their expressions are exactly the same, while only with different incoherent relaxation rate $\Gamma^\text{r}_\theta$ and the dephasing rate $\Gamma_\theta$. Specifically, if we assume the temperature felt by the QB in the BTZ spacetime and AdS spacetime is the same, then we can find the relation 
of the incoherent relaxation rate (the dephasing rate) in these two different spacetimes is given by 
\begin{eqnarray}\label{relation-BH-AdS1}
\nonumber 
\Gamma^\text{r}_{0\text{(BTZ)}}&=&\Gamma^\text{r}_{0\text{(AdS)}}+\frac{\mu^2\Delta^2}{\Omega^2}\sum^{\infty}_{n=1}
\bigg[P_{-\frac{1}{2}+i\frac{\Omega}{2\pi T}}(\cosh\alpha_n)-\zeta P_{-\frac{1}{2}+i\frac{\Omega}{2\pi T}}(\cosh\beta_n)\bigg],
\\          \nonumber 
\Gamma_{0\text{(BTZ)}}&=&\Gamma_{0\text{(AdS)}}+\frac{\mu^2\Delta^2}{2\Omega^2}\sum^{\infty}_{n=1}
\bigg[P_{-\frac{1}{2}+i\frac{\Omega}{2\pi T}}(\cosh\alpha_n)-\zeta P_{-\frac{1}{2}+i\frac{\Omega}{2\pi T}}(\cosh\beta_n)\bigg]
\\
&&+\frac{\mu^2A^2}{\Omega^2}\sum^\infty_{n=1}\bigg[P_{-\frac{1}{2}(\cosh\alpha_n)}-\zeta P_{-\frac{1}{2}}(\cosh\beta_n)\bigg],
\end{eqnarray}
and 
\begin{eqnarray}\label{relation-BH-AdS2}
\nonumber 
\Gamma^\text{r}_{\pi/2\text{(BTZ)}}&=&\Gamma^\text{r}_{\pi/2\text{(AdS)}}+\frac{\mu^2A^2}{\Omega^2}\sum^{\infty}_{n=1}
\bigg[P_{-\frac{1}{2}+i\frac{\Omega}{2\pi T}}(\cosh\alpha_n)-\zeta P_{-\frac{1}{2}+i\frac{\Omega}{2\pi T}}(\cosh\beta_n)\bigg],
\\               \nonumber 
\Gamma_{\pi/2\text{(BTZ)}}&=&\Gamma_{\pi/2\text{(AdS)}}+\frac{\mu^2A^2}{2\Omega^2}\sum^{\infty}_{n=1}
\bigg[P_{-\frac{1}{2}+i\frac{\Omega}{2\pi T}}(\cosh\alpha_n)-\zeta P_{-\frac{1}{2}+i\frac{\Omega}{2\pi T}}(\cosh\beta_n)\bigg]
\\
&&+\frac{\mu^2\Delta^2}{\Omega^2}\sum^\infty_{n=1}\bigg[P_{-\frac{1}{2}(\cosh\alpha_n)}-\zeta P_{-\frac{1}{2}}(\cosh\beta_n)\bigg].
\end{eqnarray}
Here the subscripts ``BTZ" and ``AdS" respectively denote the rates for the BTZ black hole case and AdS spacetime case. Therefore, 
we can conclude that although the average energies associated to the QB in both the BTZ black hole and AdS spacetime share the same expression,
the corresponding incoherent relaxation rate and dephasing rate are different. This means that the average energies associated to the QB in both the BTZ black hole and AdS spacetime share the same behavior with respect to the evolution time, however, their damping strengths are different due to different dissipation from vacuum field fluctuations in different spacetime. Indeed, as shown in \eqref{relation-BH-AdS1} and \eqref{relation-BH-AdS2}, 
the incoherent relaxation rate (and the dephasing rate) for the AdS spacetime case arises as the $n=0$ term of that for the BTZ black hole case, while 
the remaining terms ranging from $n=1, \cdots, \infty$ constitute the novel black hole effects. We note that these additional terms 
are parameterized by the black hole mass, $M$, and has been shown to be with more pronounced effects arising for smaller $M$, such as Fisher information \cite{FIBH}.

In the infinite time limit, i.e., $\tau\rightarrow\infty$, for both the decoherence and pure dephasing coupling cases in 
AdS spacetime the average energies in \eqref{DEAEAdS} and \eqref{DAEAdS} approach to 
\begin{eqnarray}
\langle E_\text{QB}(\infty)\rangle=\frac{\Delta}{2}\bigg\{1-\frac{\Delta}{\Omega}
\bigg(\frac{e^{\Omega/T}-1}{e^{\Omega/T}+1}\bigg)\bigg\}.
\end{eqnarray}
It is shown that in the infinite evolution time limit the average energies of the QB are independent of the kinds of dissipation, while depend on the local KMS temperature of the field in the AdS spacetime. 
Besides, we can find these average energies are exactly the same as that for the BTZ black hole case shown in \eqref{QBE}. Therefore, one can not tell the difference 
between the BTZ black hole and the AdS spacetime in terms of the average energy associated to the QB in the infinite evolution time limit.

\section{Discussions and Conclusions} \label{section5}
We find that the dynamics of the average energy stored in the QB quite depends on the weight of the driving amplitude $A$ and the energy-level spacing $\Delta$.
This is because that when $A>\Delta$, the driving may plays a dominated role in the dynamical evolution of the QB, while the dissipation due to the coupling to 
the vacuum fluctuations of quantum field plays a subordinate role. Therefore, oscillatory behavior with slow decay (which may finally leads to an equilibrium state for the QB) is presented. Oscillatory behavior results from the driving, while slow amplitude decay  is the consequence of dissipation. However, when $A<\Delta$, this is opposite. the dissipation may plays a dominated role in the dynamical evolution of the QB, while the driving plays a subordinate role. Therefore, the QB will straightly evolve to its equilibrium state while accompanied by decaying small oscillation. Small oscillation is from the driving since the driving is weak, while the decay and dissipative evolution to the equilibrium state result from the dissipation. Therefore, in the charging protocol, one can adjust the strength of the driving to manipulate the charging dynamics.

In summary, we studied how charging performances of a quantum battery, modeled as a two-level system, are influenced by the presence of vacuum fluctuations of a quantum field satisfying the Dirichlet, transparent, and Neumann boundary conditions in the BTZ spacetime. 
We obtained the analytical dynamics of this QB, and analytically calculated the average energy stored in the QB. We find that when the driving amplitude is stronger/weaker than the energy-level spacing of the QB the pure dephasing dissipative coupling results in better/worse charging performance than the decoherence dissipative coupling case. We also found that compared with the closed QB case, as a result of dissipative coupling the higher local Hawking temperature helps to improve the charging performance under certain conditions. This means that it is possible for us to extract energy from vacuum fluctuations in curved spacetime via dissipation in charging protocol. We also considered how different boundary conditions for quantum field affect the charging performance. We found that the boundary condition not only quantitatively but also qualitatively affects the the QB dynamics. It is shown that the maximum average energy stored in the QB increases with the decrease of the local Hawking temperature for the Dirichlet boundary case, while it is opposite for the transparent and Neumann boundary cases. Furthermore, we also addressed the charging stability by monitoring the energy behaviour after the charging protocol has been switched off. We found that from the perspective of the charging stability, the pure dephasing dissipative coupling may lead to better/worse charging performance than the decoherence dissipative coupling case
when the driving amplitude is stronger/weaker than the energy-level spacing of the QB. This result is valid for all the three boundary condition cases.
Our study presents a general framework to investigate relaxation effects in curved spacetime, and reveals how spacetime properties and filed boundary condition affect the charging process. This in turn, from a simple quantum device for the production, storage, and transfer of energy, may shed light on the exploration of the spacetime properties and thermodynamics via the charging protocol.

In our model, the two-level system which is coupled a massless scalar field in the BTZ spacetime can be considered as a detector. From the 
dynamics of this two-level system, we can read out the information of quantum field in the BTZ spacetime, e.g., the field correlation function \cite{PhysRevD.98.105011}, and the spacetime curvature \cite{PhysRevD.105.125011}. According to the AdS/CFT correspondence theory, actually one can find the corresponding operators in the boundary CFT \cite{PhysRevD.73.086003, hamilton2007local}. In this sense, through appropriate operation on the two-level system, we can obtain the spacetime properties and quantum field properties 
in BTZ spacetime from the two-level system's observable, thus obtain the spacetime thermodynamics, singularity, and so on. When applying the AdS/CFT correspondence theory, we can also reveal the corresponding physics in terms of a dual field theory.

Note that our QB model is general, which can be extended to different scenarios where the QB moves along different trajectories in various spacetimes. For example, we can consider the relevant physics of QB in quantum BTZ black hole \cite{QBTZ, universe10090358}, that incorporates the exact backreaction from the fields and whose thermodynamics has been explored recently \cite{PhysRevLett.130.161501, PhysRevD.109.124040, PhysRevD.110.106004, mansoori2024, PhysRevD.110.024054, PhysRevLett.133.181501}. What we need to do in the extension is to replace the Wightman function of the field shown in \eqref{BTZ-WF} with the one in the quantum BTZ black hole. Then we repeat the relevant calculations and could analyze how the parameters (or properties) of quantum BTZ black hole affect the charging performances of the QB, as done for that in BTZ black hole. Which in turn could also shed light on new properties of quantum BTZ black hole when comparing the relevant results with that for BTZ black hole case. 
When considering the backreaction into the spacetime geometry that from the quantum field, there is reason to expect that some new phenomena might emerge in the charging protocol \cite{QBTZ, universe10090358, PhysRevLett.130.161501, PhysRevD.109.124040, PhysRevD.110.106004, mansoori2024, PhysRevD.110.024054, PhysRevLett.133.181501}. Which are left for the future investigation.

\begin{acknowledgments}
This work was supported by the scientific research start-up funds of Hangzhou Normal University: 4245C50224204016; 
XL acknowledges the support by the National Natural Science Foundation of China (NSFC) under Grant No. 12065016 and the Discipline-Team of Liupanshui Normal University of China under Grant No. LPSSY2023XKTD11; JJ was supported by NSFC under Grant No. 12035005. 
\end{acknowledgments}
\bibliography{ref}
\bibliographystyle{JHEP}

\end{document}